\documentclass[lettersize,journal]{IEEEtran}
\usepackage{amsmath,amsfonts}
\usepackage{algorithmic}
\usepackage{algorithm}
\usepackage{array}
\usepackage[font=scriptsize,labelfont=sf,textfont=sf]{subcaption}
\usepackage{textcomp}
\usepackage{stfloats}
\usepackage{url}
\usepackage{verbatim}
\usepackage{graphicx}
\usepackage{cite}

\usepackage{tabularx}
\usepackage{makecell}
\usepackage[commandnameprefix=ifneeded]{changes}

\usepackage{multirow}
\usepackage{booktabs}
\usepackage{enumitem}

\usepackage[colorlinks=true,linkcolor=black,citecolor=black,urlcolor=blue]{hyperref}

\begin{document}

\title{Externalizing Hand Grip Pressure Through Real-Time Visual Feedback}

\author{Andrew L. Huard~\IEEEmembership{(Member,~IEEE)}, Avinash Ajit Nargund~\IEEEmembership{(Member,~IEEE)}, and Misha Sra
        % <-this % stops a space

%\author{IEEE Publication Technology,~\IEEEmembership{Staff,~IEEE,}
        % <-this % stops a space
\thanks{The authors are with the Department of Electrical and Computer Engineering, University of California, Santa Barbara, Santa Barbara, CA 93106 USA}% <-this % stops a space
\thanks{Manuscript received August 10, 2026; revised October 16, 2026.}}

\markboth{IEEE Transactions on Visualization and Computer Graphics,~Vol.~32, No.~10, October~2026}{IEEE Transactions on Visualization and Computer Graphics,~Vol.~32, No.~10, October~2026}%

\IEEEpubid{0000--0000/00\$00.00~\copyright~2026 IEEE}

\maketitle

\begin{abstract}
Grip pressure, including contact force and its distribution across the hand, is central to skilled grasping but is not directly visible. This makes it difficult to teach, learn, and self-correct in grip-dependent motor tasks, where verbal instruction and tactile demonstration provide only partial guidance and no real-time, spatially resolved feedback tied to the learner’s own grip. Using an instrumented glove, we compare pressure feedback visualizations for such guidance, evaluating how visualization type and, secondarily, sensor-region mapping shape pressure-matching performance in novices. In a study using baseball and table tennis as representative grip tasks (N=28), we find that a continuous gradient visualization produces significantly more accurate pressure matches and lower gaze entropy than a discrete bar display, despite participants perceiving the bar as more precise for pressure magnitude. Perceived precision and objective performance therefore diverged: the encoding participants judged more precise was not the one they matched more accurately with. Grip-tailored sensor-region mappings further reduced perceived temporal demand and effort, although this subjective workload advantage did not extend to objective matching accuracy. Overall, continuous pressure feedback supported more accurate grip matching and more concentrated viewing, and we present design considerations for real-time pressure feedback in motor skill guidance.
\end{abstract}

\begin{IEEEkeywords}
Computer aided instruction, data gloves, data visualization, gaze tracking, grasping, force measurement, sports, visual perception.
\end{IEEEkeywords}

\begin{figure*}[!t]
  \centering
  \includegraphics[width=\textwidth]{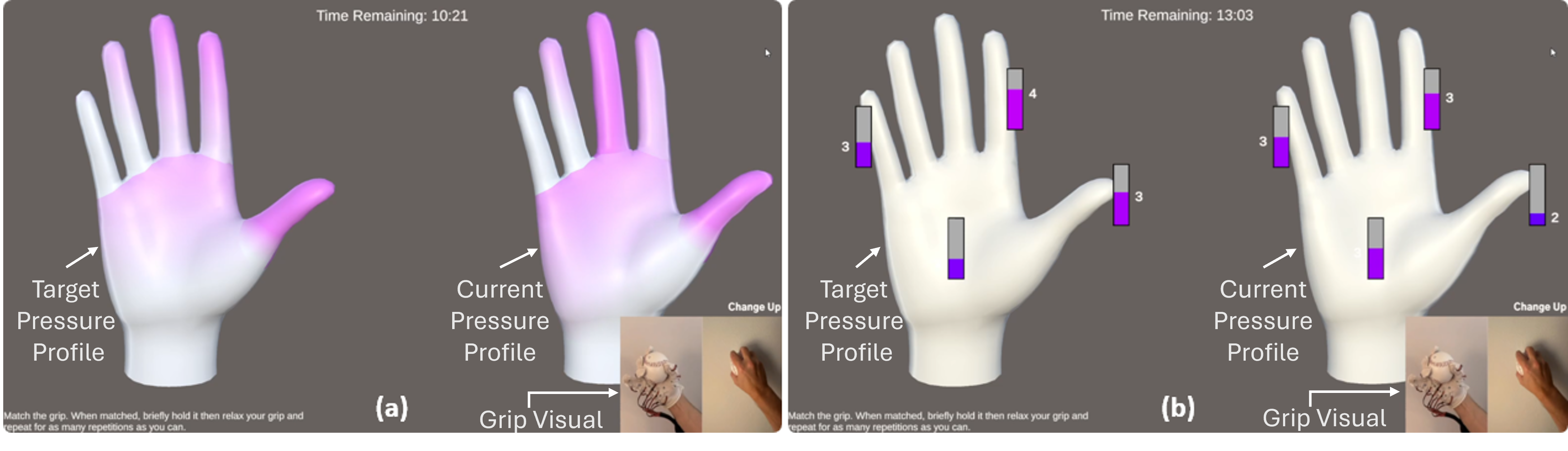}
  \caption{Visual encodings implemented in the pressure glove real-time visual feedback interface. (a) Continuous linear-vector gradient visualization providing feedback spatially aligned with the hand's sensor-region mapping. (b) Discrete pressure-sensitive bars displaying averaged regional grasp forces featuring numerical values (1--5) and color-coded ranges.}
  \label{fig:vis}
\end{figure*}

\section{Introduction}

\IEEEPARstart{H}{ow} we grasp and manipulate objects is governed by subtle, continuously adjusted grip pressure. This pressure reflects tacit, embodied knowledge acquired through repeated experience, making it difficult to quantify, observe, and communicate. Unlike gross hand motion or posture, grip pressure involves high-dimensional information about where the hand contacts an object and how force is distributed across the hand over time. These contact points and force distributions can shift continuously during a task, yet they are rarely available to either the learner or instructor as explicit feedback. Grip pressure is therefore usually inferred indirectly from cues such as hand posture, forearm tension, facial expression, and object deformation or movement.  

Acquiring grip-dependent motor skills therefore benefits from real-time, \textit{concurrent} feedback, where relevant corrections can be sensed and applied while the grip is being performed \cite{Sigrist2013a}. The channels commonly used to teach grip provide only partial support for this process. Instructional videos, books, and physical demonstrations are largely \textit{terminal}, conveying general principles or finished postures rather than feedback during the learner's own attempt \cite{Salmoni1984, Sigrist2013a}. In-person verbal cues, such as ``grip lightly,'' can be delivered in the moment, but they remain sparse and high-level, with little indication of where force should change or by how much. As a result, existing instructional channels do not provide the real-time, spatially resolved feedback needed to communicate the nuances of grip pressure.

Concurrent visual representations offer a potential solution by transforming otherwise invisible aspects of physical performance into clear, actionable feedback \cite{Kim2023Visualizing, Zhu2022MuscleRehab, Liu2025PiaMuscle, Sariya2024HandGrip}. When updated in real time, such cues can serve as instructional scaffolds, helping learners perceive deviations in their performance and correct them during the task~\cite{Salmoni1984, Delamare2016Designing}. Over time, this feedback may support the development of internal models that allow learners to interpret grip pressure and execute the skill without guidance \cite{wolpert1995internal, wolpert2011principles, Sigrist2013a}. However, because the feedback must be interpreted while the learner is actively gripping and manipulating an object, its effectiveness depends on whether the visual representation can make pressure information actionable without exceeding the learner’s limited attentional capacity~\cite{Sigrist2013a}.

\IEEEpubidadjcol % Needed somewhere in the second column on page 1 to add vspace to the footer for the publication ID

In this work, we present an exploratory investigation of how pressure data visualizations influence the perception and interpretation of tacit grip guidance. We developed a pressure-sensing glove based on the design by Sariya et al.~\cite{Sariya2024HandGrip}, instrumented with 20 force-sensitive resistors (FSRs). Rather than visualizing each sensor independently, we aggregate pressure data into biomechanically coherent regions. This approach builds on recent work in adaptive grasp interfaces~\cite{Caetano2025GraspR} and reflects the fact that the hand's many degrees of freedom are not controlled independently: joints within a finger are coupled through shared musculature~\cite{Santello1998, SchieberSantello2004}, and pressing with one digit can involuntarily recruit force from the others~\cite{Zatsiorsky2000, SchieberSantello2004}. Digits that serve a common functional role can therefore behave as a single \textit{virtual finger}~\cite{Iberall1987}, motivating the aggregation of dense per-sensor data into a smaller number of functionally meaningful sensor-regions.

Using the GRASP taxonomy, we selected two target grips whose \textit{virtual finger} structures recur across skilled domains (Table~\ref{tab:grasp-skill}): a spherical grasp, contextualized in baseball pitching but shared with other throwing and ball-handling skills, and an index-finger-extension grip, contextualized in table tennis but also relevant to golf and index-guided tool control in carpentry~\cite{Feix2016, Zhao2025EmbeddingTouch}. These sport contexts gave participants a familiar object to hold and a reason to hold it in a particular way. Target pressure profiles were recorded by having experts in baseball and table tennis execute each canonical grip on the corresponding physical object while wearing the same glove, so that novices matched a demonstrated pressure distribution rather than a numerically authored target. From these grips, we generated grip-tailored sensor-region mappings and compared them with a non-tailored generic map. We also designed two representations of the aggregated pressure data, a continuous \textit{Gradient} visualization and a discrete \textit{Bar} visualization, to examine how visual encoding affects performance and attention. We evaluated the effects of both sensor-region mapping and visual encoding on task load, attention, and pressure-matching performance. Results are averaged across the two grip types so that these comparisons are not specific to either grip.

\begin{table}[t]
  \caption{The two studied GRASP-taxonomy grips~\cite{Feix2016} and other skills sharing their virtual-finger structure.}
  \label{tab:grasp-skill}
  \centering
  \renewcommand{\arraystretch}{1.2}
  \begin{tabularx}{\columnwidth}{>{\raggedright\arraybackslash}p{0.22\columnwidth}l>{\raggedright\arraybackslash}X}
    \toprule
    \textbf{Grip} & \textbf{Task} & \textbf{Other skills} \\
    \midrule
    Precision Sphere (Grasp 13) & Baseball & Water polo throw, clay shaping, changing a lightbulb, holding a tennis ball before a serve \\
    \addlinespace
    Index Finger Extension (Grasp 17) & Table tennis & Golf, wood chisel control in carpentry, scalpel\slash electrocautery pen guidance, chef's knife with index along the spine \\
    \bottomrule
  \end{tabularx}
\end{table}

We evaluated these factors in a mixed-design user study with novice participants (N=28). Using performance metrics, surveys, gaze behavior, and qualitative feedback, we examine how visual encoding and sensor-region mapping shape a user's ability to interpret visual pressure feedback and reproduce target grip-pressure distributions. Our results show that the continuous \textit{Gradient} encoding produces significantly more accurate pressure matching and lower gaze entropy than the discrete \textit{Bar} encoding, although participants perceived the \textit{Bar} as more precise.

\begin{itemize}
    \item \textbf{RQ1:} How does pressure visualization encoding (continuous vs. discrete) affect (a) grip pressure matching performance, (b) user workload, and (c) visual attention? 
    \item \textbf{RQ2:} How does the sensor-region mapping (tailored vs. generic) affect (a) grip pressure matching performance, (b) user workload, and (c) visual attention?
    \item \textbf{RQ3:} How do users perceive each encoding, and how does each support familiarization with the feedback?
\end{itemize}

The key contributions of this work are as follows:

\begin{itemize}
    \item We present a pressure feedback system that aggregates dense hand-pressure data into task-relevant regions and visualizes it through two encodings: a continuous \textit{Gradient} visualization and a discrete \textit{Bar} visualization, evaluated against reference grip-pressure profiles recorded from experts performing each canonical grip on the physical object.
    \item Through a mixed-design user study (N=28), we show that the continuous \textit{Gradient} encoding enables significantly more accurate pressure matching and lower gaze entropy than the discrete \textit{Bar} encoding, despite participants not strongly preferring it.
    \item We found that grip-tailored sensor-region mappings reduce perceived effort and temporal demand, but we found no significant difference in objective matching accuracy relative to a generic mapping.
    \item Based on these findings, we present design implications for future grip guidance systems that externalize tacit pressure information through real-time visual feedback.
\end{itemize}

\begin{figure*}[!t]
  \centering
  \includegraphics[width=0.95\textwidth]{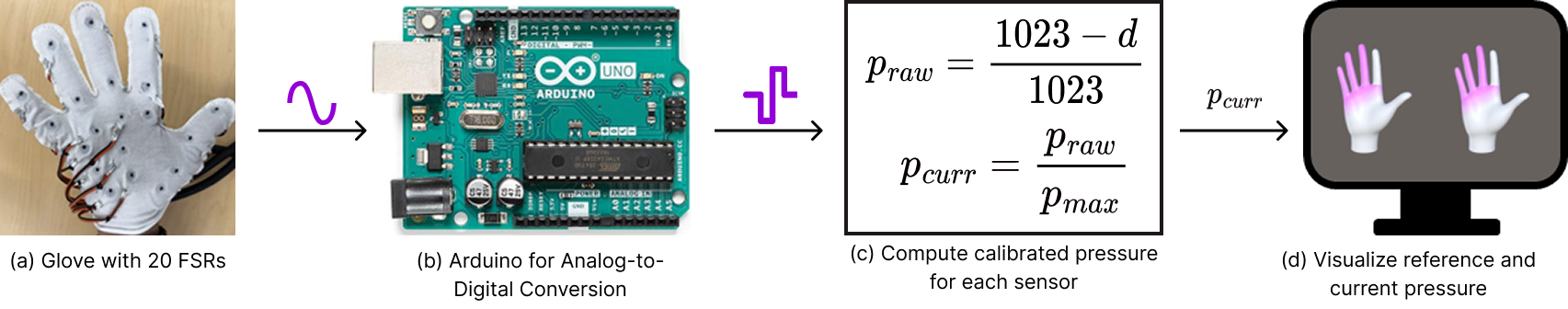}
  \caption{Pressure data is collected and visualized by (a) sampling 20 glove-mounted FSRs, (b) performing analog-to-digital conversion via an Arduino Uno R3, (c) normalizing raw pressure signals to the MVC, and (d) visualizing the reference and current pressure profiles.}
  %\Description{The pipeline is illustrated as a block diagram flowing from left to right. The first block on the left starts with a glove consisting of 20 force sensing resistors, which is denoted as block "a". An arrow with a sinusoidal symbol representing analog data then passes to the next block, which is a picture of an Arduino Uno R3, denoted as block "b". An arrow with a square-wave symbol representing digital data then passes to the next block, denoted as block "c", which lists a set of equations calculating raw pressure values from analog-to-digital converter raw data values and calculating the current pressure from the ratio of the raw pressure in the numerator to the maximum pressure in the denominator. The raw pressure values are calculated by a ratio of 1023 minus the analog-to-digital converter raw data value, represented by the letter "d", all in the numerator, followed by 1023 in the denominator. An arrow then shows the current pressure passing to the next block, which is a representation of the visualized reference and current pressure profiles, denoted as block "d"}
  \label{fig:sys_block_diagram}
\end{figure*}

\section{Related Work}

\subsection{Interactive Feedback for Motor Skill Guidance}

The sensing modality underlying a motor feedback system determines what aspects of performance can be observed, evaluated, and communicated for skill guidance. Much prior work has focused on kinematic tracking, in which a novice's pose or movement is compared against an expert reference and deviations are communicated through visual or haptic feedback~\cite{Anderson2013YouMove, Sodhi2012LightGuide, Wang2017Augmented, Chinthammit2014Ghostman}. This approach is effective for tasks where motion alignment is the primary guidance objective, but kinematic data provides limited access to the internal or contact-level mechanics that produce the observed motion.

To address this limitation, recent feedback systems have incorporated neuromuscular sensing, such as surface electromyography (sEMG) and electrical impedance tomography (EIT), to visualize hidden muscle engagement~\cite{Zhu2022MuscleRehab, Liu2025PiaMuscle, Kim2023Visualizing}. These approaches extend feedback beyond external pose or movement to the physiological activity driving action. However, neither kinematic nor neuromuscular sensing directly captures how force is distributed across the contact surface between the hand and object. In grip-dependent domains such as sports, musical performance, and tool use, expert execution depends not only on movement trajectory or muscle activation, but also on where and how much pressure is applied over time. Despite its relevance to skilled performance, localized contact force distribution remains underexplored in interactive motor guidance systems. In this work, we investigate real-time visual feedback for grip pressure as a means of externalizing this otherwise difficult-to-observe dimension of skilled action.

\subsection{Feedback Modalities in Motor Guidance}

Motor guidance systems also differ in how they deliver feedback and in how much responsibility they assume for the user's action. Prior work has explored feedback through sound~\cite{Masai2022Sonification}, visual displays~\cite{Liu2025PiaMuscle, Sariya2024HandGrip}, and haptic cues~\cite{Lee25}. Haptic guidance systems can intervene directly in physical execution through physical guidance, force feedback, impedance control, exoskeletons, or electrical muscle stimulation~\cite{Langerak20b, Shim20, Tsai22, Hashimoto22, Lee25, Lopes2016MusclePlotter}. By partially imposing or constraining the movement, these systems can reduce early errors and physically guide users toward a target action~\cite{Seim2014PassiveHaptic, Lee25, Langerak20b}.

However, physically enforced guidance can also reduce the need for users to actively interpret errors and generate corrective motor commands themselves. Prior work on motor learning has noted that excessive or overly continuous guidance can produce dependency on the external scaffold and limit learning transfer~\cite{Sigrist2013a, Pezent2018SeparatingHaptic, Salmoni1984}. Techniques such as bandwidth feedback address this issue by triggering assistance only when error exceeds a predefined threshold~\cite{Sherwood1988}. In contrast, concurrent visual feedback supports a different form of guidance that informs rather than enforces. Users must perceive the feedback, interpret their state relative to a target, and intentionally adjust their own movement in a closed-loop process~\cite{Lebrun2025MentorGuided, Sigrist2013a}. Our approach follows this active-correction model. Rather than physically guiding the hand, we provide real-time visual access to localized grip pressure so that users can regulate their own contact force while performing the task.

\subsection{Visualizing Force for Closed-Loop Control}

Visual representations of embodied data, such as heatmaps, contours, spatial overlays, and abstract displays, can translate otherwise hidden physiological or contact-level signals into actionable feedback~\cite{Zhu2022MuscleRehab, Liu2025PiaMuscle, Sodhi2012LightGuide, Sariya2024HandGrip, Anderson2013YouMove}. For grip pressure guidance, however, the feedback must be interpreted while the user is actively grasping and manipulating an object. This creates a design challenge such that detailed feedback may provide more information, but can also exceed the learner's attentional capacity during concurrent motor control~\cite{Sariya2024HandGrip, Sigrist2013a}. Effective representations must therefore balance precision, interpretability, responsiveness, and cognitive load~\cite{Sweller2010, Endsley95, wickens2008multiple}. They must also preserve enough spatial structure for users to understand where pressure should change, without overwhelming them with dense sensor-level data.

The choice of visual encoding shapes the perceptual task users must perform. Position- and length-based encodings, such as bars, support accurate readout of individual magnitudes~\cite{ClevelandMcGill1984}, whereas continuous spatial fields can be read more holistically, allowing users to extract summary information from distributed features rapidly and with limited attention~\cite{Ariely2001, AlvarezOliva2009, Whitney2018}. The spatial organization of the feedback also matters. Feedback placed on or near the hand region it describes can support more direct action, whereas feedback shown apart from the hand must first be mentally mapped back to the relevant contact area~\cite{WickensCarswell1995}. These considerations suggest a tradeoff between discrete displays optimized for reading specific pressure values and continuous displays optimized for spatial matching and rapid interpretation. They also motivate reducing dense sensor data into meaningful regions rather than presenting each pressure sensor independently. Our study examines these design choices by comparing continuous and discrete visual encodings, as well as grip-tailored and generic sensor-region mappings, for real-time grip pressure matching.

\section{System Design} \label{sys:hw}

To capture and externalize tacit grip forces, we developed a real-time visual feedback system consisting of a sensor-equipped wearable glove and a custom visualization application. This section describes the hardware used to capture contact force, the design rationale for aggregating sensor data into grip-relevant regions, and the visual encodings used to provide real-time feedback.

\subsection{Pilot Study} \label{hw:pilot}

We conducted a pilot study to determine how pressure data from the glove should be aggregated for real-time grip guidance. We compared three levels of aggregation: \textit{no aggregation}, where each sensor was shown independently; \textit{regional aggregation}, where sensors were grouped by finger or hand region; and \textit{full aggregation}, where the whole hand was represented as a single pressure value. The \textit{no aggregation} condition was consistently judged unworkable: participants could attend to only a few hand regions at once, had insufficient time to respond before the target updated, and reported higher effort without improved matching. The \textit{full aggregation} condition was easiest to control, but too coarse to represent the target grips. In contrast, \textit{regional aggregation} provided the best balance between usability and grip specificity.

The pilot also revealed a consistent observation that shaped our final design: participants experienced strong biomechanical coupling between hand regions. They noted that pressure applied to one part of a finger altered readings elsewhere on the same finger, and that adjusting one region could unintentionally perturb others, such as the finger and palm regions involved in stabilizing the same object. These findings motivated our final sensor-region mapping strategy, in which sensors were allocated into four task-relevant sensor-regions based on the functional \textit{virtual finger} structure of each target grasp~\cite{Feix2016, Iberall1987} and prior evidence of inter-finger coupling during force production~\cite{Zatsiorsky2000}. This design aimed to preserve task-relevant pressure structure while reducing attentional and control demands.

\subsection{Pressure Glove Design} \label{hw:sys}

To provide visual feedback of contact force distribution, we implemented a pressure-sensing glove based on the design by Sariya et al.~\cite{Sariya2024HandGrip}. The glove integrates 20 force-sensitive resistors as shown in Figure~\ref{fig:hand_sensors} (FSRs; DF9-16\footnote{DF9 Series Datasheet: \url{cdn.velleman.eu/downloads/25/infosheets/fsr_datasheet.pdf}}) into a cotton inspection glove\footnote{Cotton inspection glove, Amazon ASIN: B08NB72PRT} to capture pressure across the hand. Raw analog signals are routed through CD74HC4067 multiplexers\footnote{CD74HCx4067 Datasheet: \url{ti.com/lit/ds/symlink/cd74hc4067.pdf}} to an Arduino Uno R3, which digitizes the 20-sensor array and transmits the data over serial to a custom Unity desktop application (v6000.2.6f2) at 22 Hz. The Unity application renders real-time feedback on a hand mesh at 30 FPS, holding the most recent sensor telemetry between updates, as detailed in Section~\ref{sys:viz}.

\subsection{Sensor-Region Mapping} \label{sys:zone}

\begin{figure}[!b]
  \centering
  \includegraphics[width=0.45\textwidth]{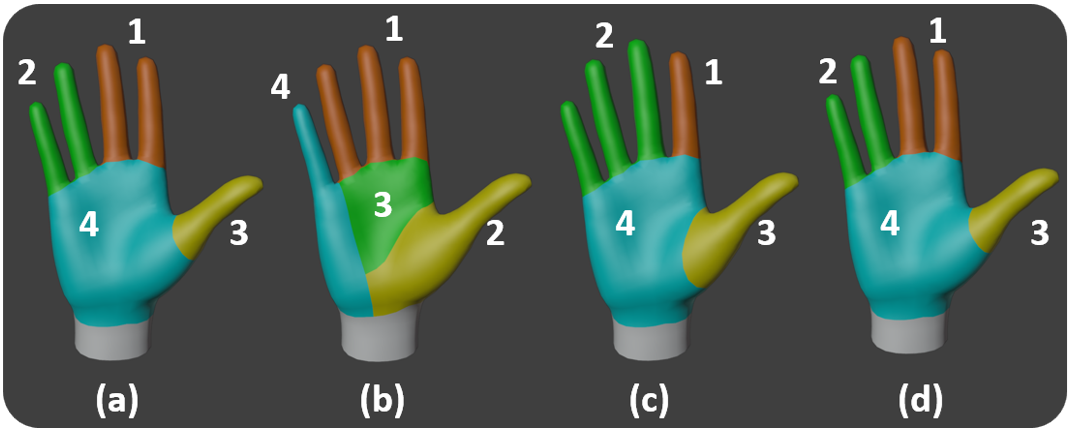}
  \caption{Mapping of hand segmentation into four aggregate sensor-regions. For baseball, (a) represents Sensor-Region Mapping A while (b) represents Sensor-Region Mapping B. For table tennis, (c) represents Sensor-Region Mapping A while (d) represents Sensor-Region Mapping B.}
  %\Description{Four sensor pooling strategies are shown, two for each sport, with zone maps illustrated across the 3D mesh of the hand. Starting from left to right, the first represents Zone Map A for baseball, the second represents Zone Map B for baseball, the third represents Zone Map A for ping pong, while the fourth represents Zone Map B for ping pong.}
  \label{fig:zone_map}
\end{figure}

\begin{figure}[!t]
  \centering
  \includegraphics[width=0.3\textwidth]{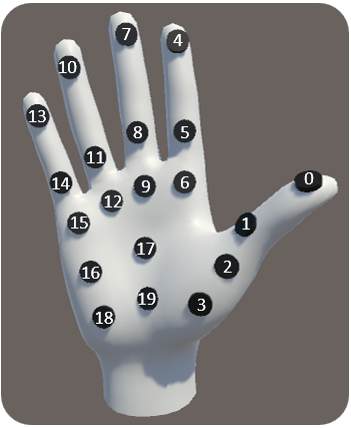}
  \caption{Location of 20 FSR sensors distributed across the hand. Each FSR component is denoted by a black circle and its corresponding sensor ID number (0-19).}
  %\Description{An image of a hand avatar is shown with the location of 20 FSR sensors illustrated across the hand. Sensor 0 is shown on the tip of the thumb, sensor 1 is shown on the thumb proximal, sensor 2 is shown on the thumb metacarpal, sensor 3 is shown on the thumb base, sensor 4 is shown on the index finger tip, sensor 5 is shown on the index finger proximal, sensor 6 is shown on the index finger metacarpal, sensor 7 is shown on the middle finger tip, sensor 8 is shown on the middle finger proximal, sensor 9 is shown on the middle finger metacarpal, sensor 10 is shown on the ring finger tip, sensor 11 is shown on the ring finger proximal, sensor 12 is shown on the ring finger metacarpal, sensor 13 is shown on the pinky finger tip, sensor 14 is shown on the pinky finger proximal, sensor 15 is shown on the pinky finger metacarpal, sensor 16 is shown on the middle left of the palm, sensor 17 is shown on the center of the palm, sensor 18 is shown on the bottom left of the palm, and sensor 19 is shown on the bottom center of the palm.}
  \label{fig:hand_sensors}
\end{figure}

The pilot study established that displaying all 20 sensors independently (Figure~\ref{fig:hand_sensors}) was unworkable, while full-hand aggregation was too coarse to represent task-relevant grip structure. An intermediate regional granularity was therefore most usable, but the pilot did not specify which sensors should be mapped together into a region. This outcome is consistent with prior work showing that reducing the complexity of concurrent visual feedback can support early performance by lowering attentional demand~\cite{Sariya2024HandGrip, Delamare2016Designing}. We therefore designed sensor-region mappings that aggregate dense pressure data into a small number of task-relevant regions across the hand (Figure~\ref{fig:zone_map}).

To define these sensor-region mappings, we used the GRASP taxonomy~\cite{Feix2016}, which characterizes grasps partly in terms of \textit{virtual fingers}: functional units in which multiple digits act together during grasping. This provided a biomechanical basis for mapping sensors into regions according to the functional structure of each target grip rather than treating each sensor independently. To isolate the effect of sensor-region mapping from the amount of information shown, all study conditions used four regions; the tailored and generic mappings differed in how sensors were assigned to those regions, not in the number of regions presented.

The table tennis tasks (forehand, backhand, and slice) use the handshake grip, classified in GRASP as Index Finger Extension (Grasp 17), in which the index finger operates independently of the other digits. This grip structure maps onto four task-relevant sensor-mapping regions: the palm, the middle-ring-pinky region, the index finger, and the thumb. The baseball tasks (fastball, change-up, and tag-out) use a spherical grip, classified as Precision Sphere (Grasp 13), which decomposes into the palm, the thumb, and the remaining four fingers. Because pitch control relies strongly on the index and middle fingers, we further divided the four-finger sensor-region into index-middle and ring-pinky regions. These VF-informed, task-specific assignments define the \textit{grip-tailored sensor-region mapping} (Mapping A; see Figure~\ref{fig:zone_map}).

The \textit{generic sensor-region mapping} (Mapping B) preserves the same four-region structure but shifts the boundary between finger groups so that the mapping no longer follows the task-specific functional organization of the grip. For table tennis, the index finger is grouped with the middle finger rather than treated independently. For baseball, the ring finger is grouped with the index-middle region rather than the pinky. Thus, Mapping B is not a single mapping reused across sports; rather, it is a non-tailored comparison mapping defined relative to each sport's tailored mapping. This design allows us to compare whether aligning sensor regions with task-specific grip structure improves workload, attention, and pressure-matching performance while holding the number of displayed sensor-regions constant.

\begin{figure*}[!t]
  \centering
  \includegraphics[width=0.95\textwidth]{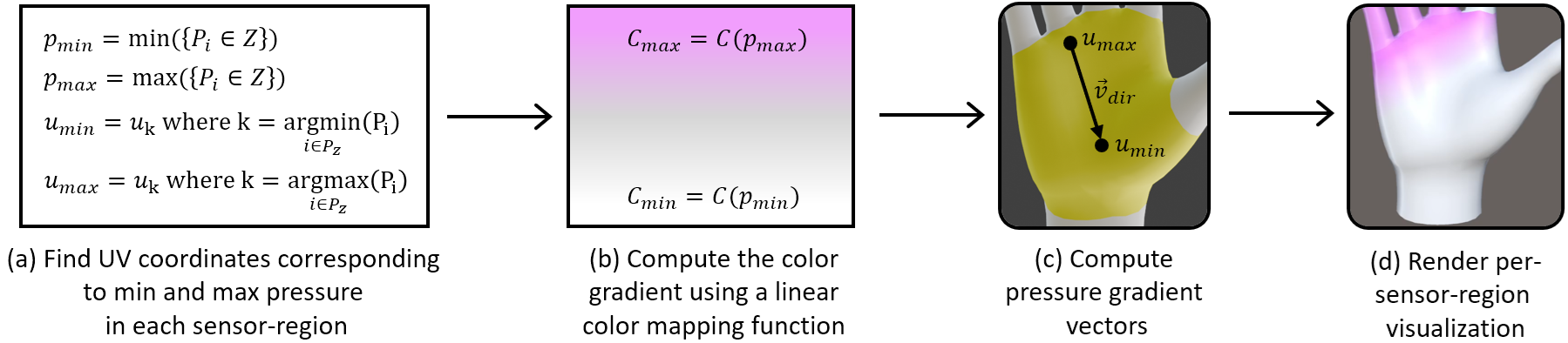}
  \caption{To visualize the gradient, we (a) aggregate the extrema of the calibrated pressure values and their UV coordinates, (b) define the color range via a linear mapping function, (c) calculate the UV vector direction, and (d) apply the resulting linear interpolation across each mesh sensor-region.}
  %\Description{A four-panel technical diagram illustrating the process of visualizing a pressure gradient on a 3D mesh. Panel (a) shows computation of pressure extrema (maximum and minimum values) and their corresponding UV coordinates. Panel (b)  displays a color gradient representing a linear mapping function translates numerical pressure values into a specific color. Panel (c) illustrates the calculation of the UV vector direction, shown as directional arrows overlaid on the mesh to indicate the flow of the gradient. Panel (d) shows the final result, where a smooth linear interpolation of colors is applied across individual mesh zones, creating a continuous visual gradient based on the previously defined parameters.}
  \label{fig:gradient_vis_block_diagram}
\end{figure*}

\subsection{Visualizations} \label{sys:viz}

To investigate how visual encoding affects real-time grip matching, we implemented two representations of the same aggregated sensor-region data: \textit{Gradient}, which renders zonal pressure as a spatially aligned continuous color field, and \textit{Bar}, which represents average zonal pressure as discrete bars on a $1-5$ numerical scale.

\subsubsection{Gradient Visualization}

The \textit{Gradient} visualization projects pressure values onto the corresponding regions of a 3D mesh via a real-time spatial interpolation pipeline (Figure \ref{fig:gradient_vis_block_diagram}). The mesh is the rigged hand model from the Meta Interaction SDK\footnote{\url{https://developers.meta.com/horizon/documentation/unity/unity-isdk-interaction-sdk-overview/}} (the stock \texttt{OculusHand} asset), rendered at a fixed scale for all participants; it was not resized to individual hand dimensions.

For each region, sensor locations on the mesh define control points in UV space. At each update, the system identifies the sensors with the minimum and maximum normalized pressure values in that sensor-region and uses their UV coordinates $\mathbf{u}_{\min}$ and $\mathbf{u}_{\max}$, to define the gradient direction vector:

\begin{equation} \label{eq:dir}
    \mathbf{v}_{\text{dir}} = \mathbf{u}_{\max} - \mathbf{u}_{\min}
\end{equation}

If $|\mathbf{v}_{\text{dir}}|^{2} < 0.00001$, no stable spatial direction can be estimated. In this case, the sensor-region is rendered with a uniform color corresponding to the region's pressure value. Otherwise, the normalized direction vector is:

\begin{equation} \label{eq:d}
    \mathbf{\hat{d}} = \frac{\mathbf{v}_{\text{dir}}}{\|\mathbf{v}_{\text{dir}}\|}
\end{equation}

Let $\mathbf{u}$ be the UV coordinate of the current pixel being evaluated. The vector $\mathbf{v}$ from the minimum pressure node to the current pixel is:

\begin{equation} \label{eq:v}
    \mathbf{v} = \mathbf{u} - \mathbf{u}_{\min}
\end{equation}

The scalar projection of the pixel onto the normalized gradient axis determines its raw distance along the line, which is converted into a normalized interpolation factor $t \in [0, 1]$:

\begin{equation} \label{eq:t}
    t = \text{clamp}\left( \frac{\mathbf{v} \cdot \mathbf{\hat{d}}}{\|\mathbf{v}_{\text{dir}}\|}, 0, 1 \right)
\end{equation}

Finally, let $C_{\text{low}}$ and $C_{\text{high}}$ be the boundary colors corresponding to the minimum and maximum pressure values. The color $C_{\text{px}}$ for the pixel is calculated by linear interpolation:

\begin{equation} \label{eq:c}
    C_{\text{px}} = C_{\text{low}} + t(C_{\text{high}} - C_{\text{low}})
\end{equation}

\subsubsection{Bar Visualization}

The \textit{Bar} visualization displays one bar next to each of the hand's $K=4$ sensor-regions ($Z_k$). To calculate the fill level of each bar, we first compute the normalized bounded pressure, $P_{\text{norm},i}$, for each sensor $i \in Z_k$. This value is calculated using the sensor's raw pressure reading, $P_{\text{raw},i}$, and its calibrated maximum limit, $P_{\text{max},i}$:

\begin{equation} \label{eq:p}
    P_{\text{norm}, i} = \text{clamp}\left(\frac{P_{\text{raw}, i}}{P_{\text{max}, i}}, 0, 1\right)
\end{equation}

The average normalized pressure, $\bar{P}_{k}$, for a given sensor-region $k$ is then calculated as the arithmetic mean of the bounded sensor readings in that region, where $N_k$ is the number of active sensors in the region:

\begin{equation} \label{eq:pk}
    \bar{P}_{k} = \frac{1}{N_k} \sum_{i \in Z_k} P_{\text{norm}, i}
\end{equation}

The bar's visual fill percentage is driven directly by $\bar{P}_{k}$. Its numerical annotation is scaled from $1$ to $5$ to provide a discrete readout of the same aggregated pressure value.

\subsection{Data Logging}

Grip pressure data were continuously logged in the Unity application for each trial, including initial calibration values and raw analog-to-digital converter (ADC) readings from all 20 sensors. In parallel, participant gaze was recorded on a desktop computer using Beam Eye Tracker software with a Google Pixel 8a webcam positioned at a 35-degree angle. We time-synchronized the gaze and pressure streams for each trial and post-processed the combined dataset offline using the methods described in Section~\ref{sys:measures}.

\subsection{Target Pressure Profiles} \label{sec:targets}
Each sport was associated with three target grips: fastball, change-up, and tag-out for baseball, and forehand, backhand, and slice for table tennis.

The reference profile for each grip was recorded by an expert who executed the grip on the same physical object used in the study while wearing the instrumented glove. Two experts contributed profiles, one recording the three baseball grips and the other the three table tennis grips. Each held each grip for three repetitions. Every recording was normalized to the expert's MVC for the corresponding object using the same calibration procedure later applied to study participants (Section~\ref{sec:procedure}), then averaged across repetitions to yield one reference profile per grip. Because profiles were stored as a per-sensor proportion of MVC rather than as raw ADC values, the same profile could be rescaled to each participant's own grip strength; the rescaled profile supplies $P_{\mathrm{ideal},i}$ in Eq.~\ref{eq:mae}.

Throughout each trial, the target profile was rendered on a second hand mesh displayed alongside the participant's live hand, using the same visual encoding as the active condition, so that participants could compare the two directly. The target remained static for the duration of a trial.

\section{User Evaluation}

We conducted a user study to examine how visualization type and sensor-region mapping affect novice user ability to match target grip-pressure profiles using the pressure glove system described in Section~\ref{sys:hw}. Sessions were conducted in a quiet laboratory. The study ran on a desktop computer with a 27-inch monitor ($1920 \times 1080$ resolution) and an NVIDIA RTX 3070 GPU. The Unity visualization ran at 30 FPS, with pressure telemetry updated at 22 Hz as described in Section~\ref{hw:sys}. Participants were seated $0.67$ meters from the screen and rested their gloved arm on a table to stabilize the hand throughout the session. The study was approved by the University of California Santa Barbara Institutional Review Board (Protocol \#27-26-0012), and all participants provided informed consent before participation. Below, we describe the study design, tasks, procedure (Figure~\ref{fig:study_procedure}), measures, and participants.

\subsection{Study Procedure and Task} \label{sec:procedure}

\begin{figure*}[!t]
    \centering
    \includegraphics[width=0.95\linewidth]{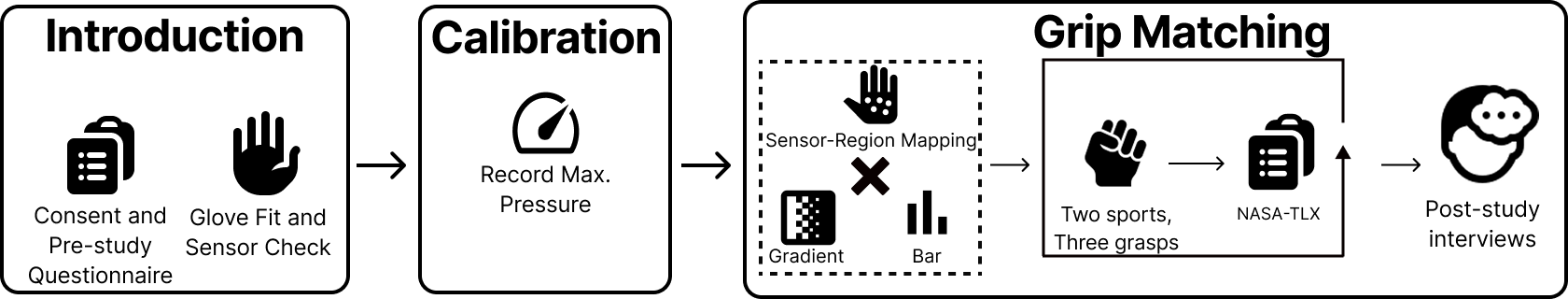}
    \caption{Study procedure. Participants provided informed consent, confirmed glove fit and sensor functionality, and completed a pre-study demographic and experience questionnaire. In the calibration phase, three MVC attempts per object (baseball and table tennis paddle) were averaged to normalize target pressure profiles to each participant's grip strength. During the practice period, participants familiarized themselves with all six grip profiles, both visualization conditions, and the grip-match-release protocol. The grip-matching phase comprised four counterbalanced blocks (balanced Latin square), each pairing one sport with one visualization. In each 20-second trial, participants matched a target pressure profile displayed on a virtual hand, maximizing successful matches. After each block, participants completed the NASA-TLX. The study concluded with a post-study preference questionnaire and a semi-structured interview.}
    %\Description{The study procedure begins in the first block on the left titled Introduction. First participants are asked to complete a consent and pre-study questionnaire, then a glove fit and sensor check are performed. After this, the second block in the diagram shows the calibration method, which records the maximum pressure. Finally, the grip matching block illustrates the main study procedure consisting of selecting a combination of sport (baseball or ping pong) and visualization condition (gradient or bar), when then defines the four trial blocks. Each trial block consists of three grips with three repetitions of each grip trial. At the end of each trial block, participants completed the NASA-TLX. At the conclusion of the study, participants completed post-study interviews.}
    \label{fig:study_procedure}
\end{figure*}

Participants began the study (Figure~\ref{fig:study_procedure}) by providing informed consent and completing a demographic questionnaire along with a pre-study survey assessing their prior experience and familiarity with wearable haptics and visual feedback in sports training through 5-point Likert scale items.

Before the grip-matching trials, participants completed a grip-pressure calibration and a practice period. During calibration, participants performed three maximum voluntary contraction (MVC) attempts per object, once for the baseball and once for the table tennis paddle, which were averaged to normalize target pressure profiles to each participant's grip strength. In the practice period, participants tried all six grip profiles across both objects and both visualization conditions, familiarizing themselves with the color encodings and the grip-match-and-release protocol to register a match.

Each trial presented one target grip for 20 seconds, during which participants adjusted their grip on the physical object to match the target pressure profile displayed on a virtual hand, completing as many successful matches as possible within that window. Trials were organized into blocks, each pairing one sport (baseball, table tennis) to one visualization (Gradient, Bar). Each block comprised nine trials, consisting of the three target grips for that sport presented three times each in randomized order. Each participant completed all four blocks, for a total of 36 trials.

The grip-matching phase used a $2 \times 2$ mixed-factorial design, with Visualization as the within-subjects factor and sensor-region mapping as the between-subjects factor. Sport was not treated as an experimental factor, and the two blocks sharing a visualization condition were averaged to yield a single value per visualization per participant (Section~\ref{sys:measures}). The order of the four blocks was counterbalanced across participants using a Balanced Latin square design. After each block, participants completed the NASA Task Load Index (TLX)~\cite{hart_development_1988} to assess subjective workload.
At the end of the study, participants indicated their preferred visualization and took part in a semi-structured interview (prompts listed in Table~\ref{tab:interview-prompts}) with the researcher.

\begin{table}[t]
\centering
\caption{Semi-structured interview prompts.}
\label{tab:interview-prompts}
\begin{tabular}{p{0.02\textwidth} p{0.42\textwidth}}
\toprule
\textbf{\#} & \textbf{Prompt} \\
\midrule
I1 & Can you describe how you used the visualization to adjust your grip? \\
I2 & What were you paying attention to the most? \\
I3 & Can you walk us through your decision making process for how you decided that you matched the template? \\
I4 & When the visualization changed, how clear was it what you should do with your hand? \\
I5 & Were adjustments obvious or did you have to reason them out? \\
I6 & How did your experience differ between the two visualizations? \\
I7 & Did one support quicker or more confident adjustments? \\
I8 & Was there a moment when the visualization didn't help or was confusing? \\
I9 & What would have made it clearer in that moment? \\
\bottomrule
\end{tabular}
\end{table}

\subsection{Participants}

Following initial pilot tests, a power analysis was conducted to determine the required sample size. Assuming medium effect size ($f = 0.25$), significance level of $\alpha = 0.05$, a statistical power of $1 - \beta = 0.8$ and correlation of $0.6$ among repeated measures with $2$ groups and $2$ measurements, the analysis indicated a target sample of 28 participants.

We subsequently recruited 28 participants aged between 18 to 37 years ($M = 23, SD = 4.6$). All participants were right-handed and reported no color vision deficiencies or any hand-related conditions that could affect grip performance. Prior experience with wearable haptics (e.g., fitness trackers) was moderate-to-high for 10 participants, while 6 reported regular use of visually-augmented sports training applications. Participants received \$20 compensation for the 50-minute session.

\subsection{Dependent Variables} \label{sys:measures}

In order to evaluate the effect of region-based sensor mapping and visualization on grip-matching performance and workload, we collected data from the pressure glove trials, including normalized pressure readings from the 20 FSR sensors, eye-gaze positions, post-block NASA Task Load Index~\cite{hart_development_1988} responses, and visualization preferences.

For both the grip-matching performance and gaze measures, per-trial values were averaged across the nine trials of each block. These block-level means, along with the NASA-TLX scores collected after each block, were then averaged across the two blocks sharing a visualization, yielding one value per participant per visualization.

\paragraph{Grip-Matching Performance}

Grip-matching accuracy was quantified per trial using Mean Absolute Error (MAE), computed as the average absolute difference across $N$ active sensors (Eq. \ref{eq:mae}) (varied by sport), with pressures normalized to [0,1] via system calibration:
\begin{equation} \label{eq:mae}
    \text{MAE(t)} = \frac{1}{N} \sum_{i=1}^{N} |P_{\text{user},i}(t) - P_{\text{ideal},i}(t)|
\end{equation}

From the MAE time series data, we derived four performance metrics based on the concept of a ``match''. A match was defined as a local minimum in the MAE signal, corresponding to the completion of a grip repetition as indicated by a full grip release between consecutive repetitions. Based on this, we computed four metrics, (1) mean MAE across matches, (2) minimum MAE across matches, (3) time to first match, and (4) total number of matches.

\paragraph{Gaze Metrics}

To objectively evaluate visual attention and scanning behavior, we computed gaze entropy, horizontal gaze dispersion, and vertical gaze dispersion over each trial. Gaze entropy serves as a proxy for attention and cognitive load, as higher entropy has been empirically shown to correlate with greater attention dispersion and lower task performance~\cite{LeeJung2025, ayala2023effects, Bhavsar2017}. We computed gaze entropy by discretizing the $1920 \times 1080$ screen into 1600 bins of size $48 \times 27$ pixels each, deriving a probability distribution of gaze positions over these bins, and applying the Shannon entropy formula to quantify the uniformity of gaze distribution.

For gaze dispersion, we quantified the spread of gaze in the horizontal and vertical directions separately. Raw gaze positions in screen coordinates were first converted to visual angles relative to the mean gaze position over the trial block, after which we computed the standard deviation of these angles along each axis to obtain horizontal and vertical dispersion measures respectively.

\section{Results}

The study followed a mixed design, with sensor-region mapping as a between-subjects factor and visualization type as a within-subjects factor. Data were collected across two representative tasks, baseball and table tennis, which served to contextualize the grips of interest (see Table~\ref{tab:grasp-skill}) within realistic sport-specific actions. Because each sensor-region mapping is derived from the grip it accompanies, grip and mapping are not independent factors, and we do not model grip separately. All metrics were therefore averaged across the two blocks sharing a visualization prior to modeling to obtain estimates that are not specific to either grip. To account for this mixed-design, we used linear mixed models (LMMs) and generalized linear mixed models (GLMMs) to evaluate statistical differences across experimental conditions. Subjective workload data (NASA-TLX) were modeled using LMMs fitted with the \texttt{lme4}~\cite{bates2015fitting} and \texttt{lmerTest}~\cite{kuznetsova2017lmertest} packages in R. Performance metrics were modeled using GLMMs fitted with the \texttt{glmmTMB} package~\cite{brooks2017glmmtmb}, with family and link function chosen based on the expected scale and distribution of each metric. The sensor-region mapping, visualization type, and their interaction were entered as fixed effects, and participant ID was included as a random intercept to account for inter-individual variability. Residual diagnostics for all models were assessed using the \texttt{DHARMa} package~\cite{hartig2016dharma}, which generates simulation-based residuals to check distributional fit, dispersion, and other misspecification. Model convergence was assessed separately via optimizer warnings, gradient values, and Hessian conditioning. We applied Type-II Wald tests using the \texttt{car} package~\cite{fox2018r} to assess the significance of fixed effects and used \texttt{emmeans}~\cite{lenth2019package} for post-hoc pairwise comparisons when interaction effects were significant.

\subsection{Grip-Matching Performance} \label{results:matching}

\begin{figure}[ht]
    \centering
    \subfloat[Number of matches\label{fig:matches}]{%
        \includegraphics[width=0.45\linewidth]{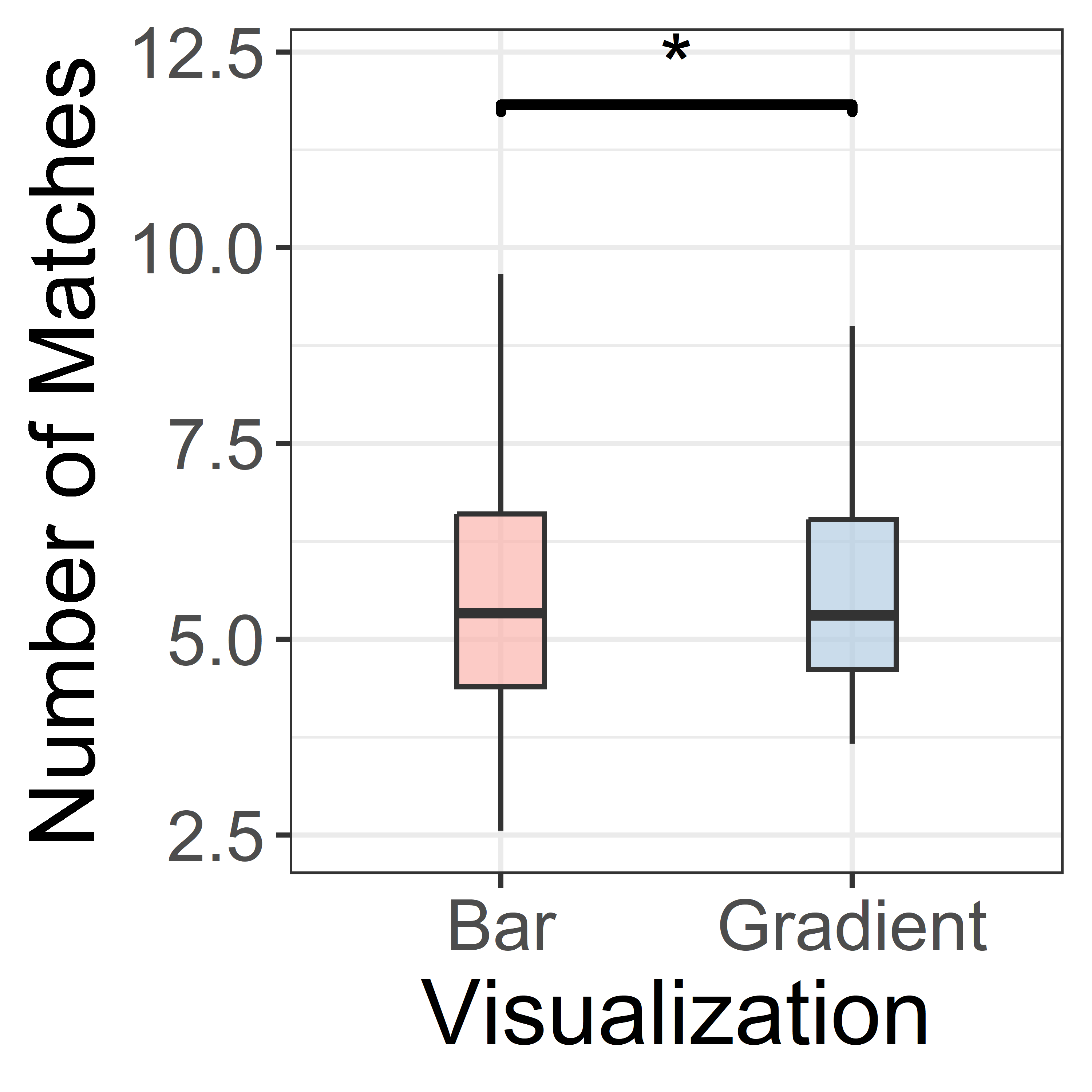}}
    \hfil
    \subfloat[Time to first match\label{fig:time_to_first_match}]{%
        \includegraphics[width=0.45\linewidth]{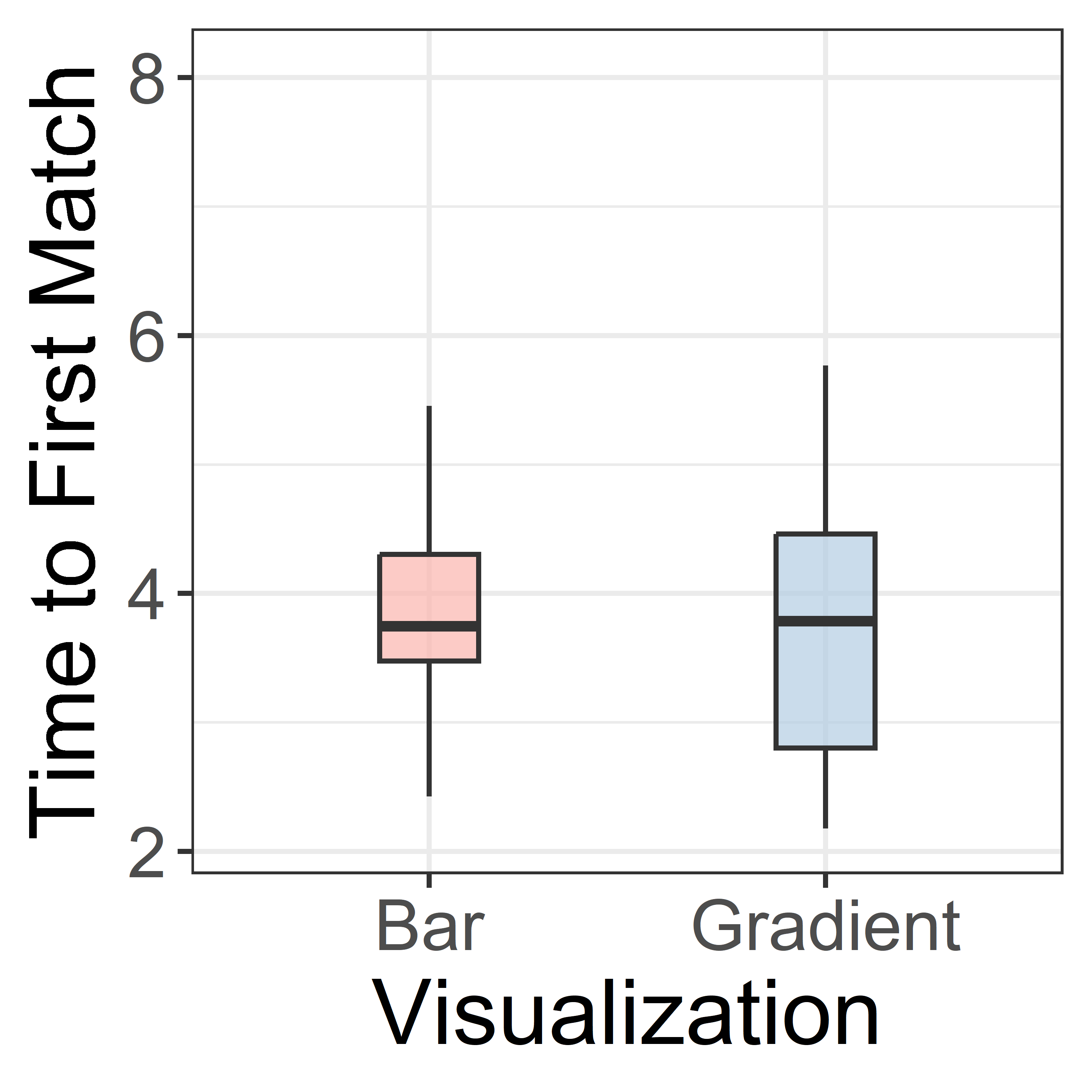}}
    \\[6pt]
    \subfloat[Avg match MAE\label{fig:avg_match_mae}]{%
        \includegraphics[width=0.45\linewidth]{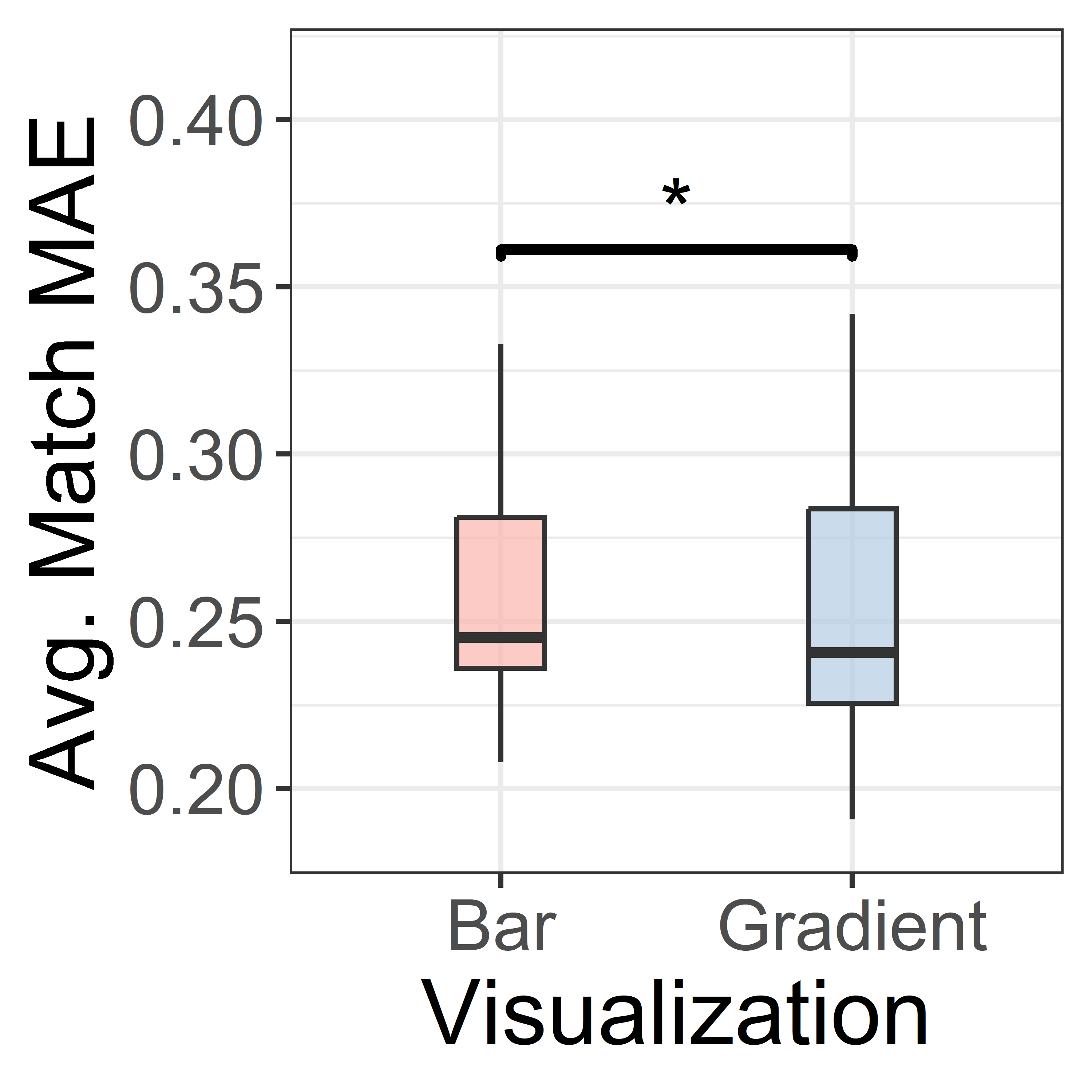}}
    \hfil
    \subfloat[Best match MAE\label{fig:best_match_mae}]{%
        \includegraphics[width=0.45\linewidth]{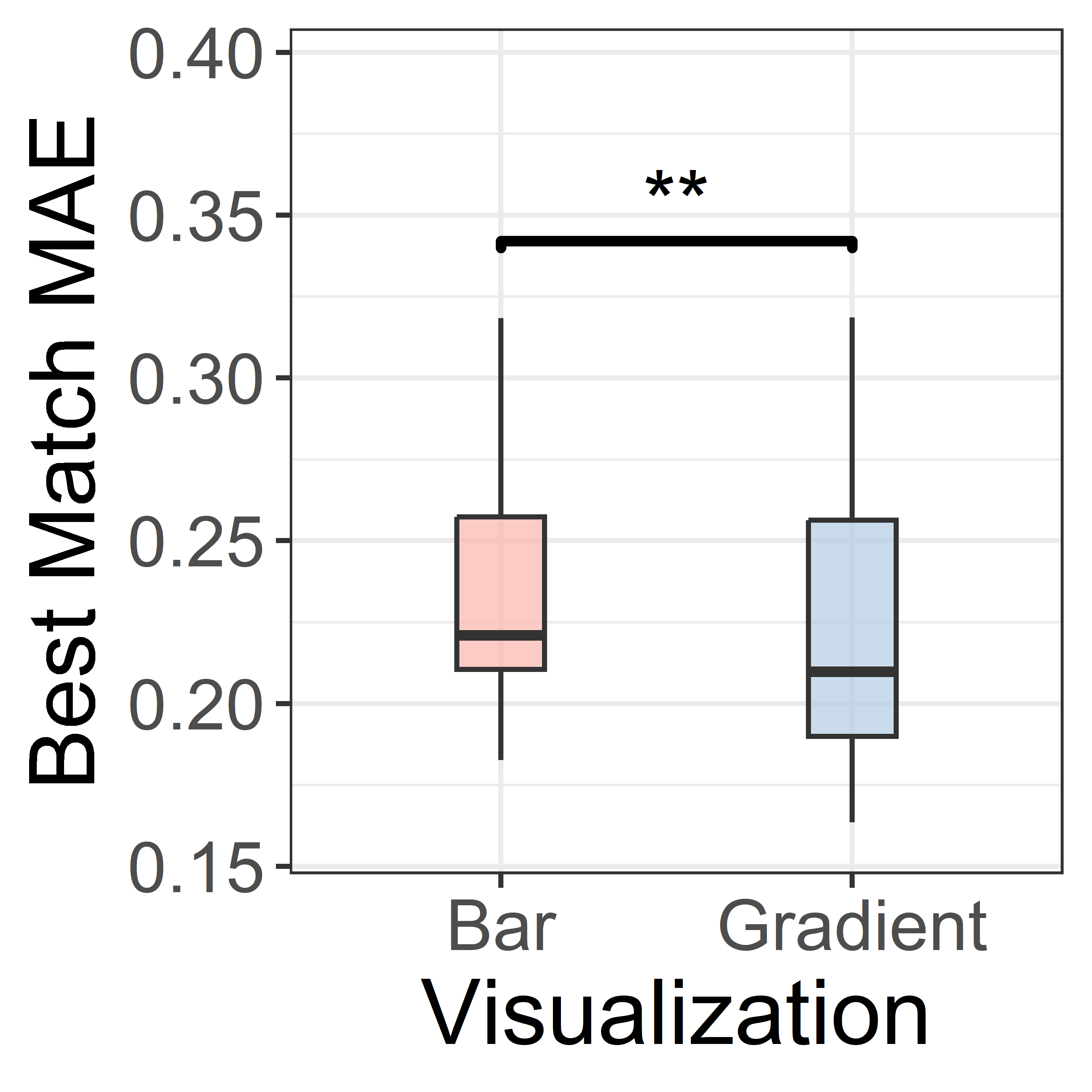}}
    \caption{The performance of the participants is measured using mean (a) number of matches, (b) time to first match, (c) average match MAE and (d) best match MAE. Participants made significantly more matches per block and were significantly more accurate with the \textit{Gradient} visualization. The time to first match did not differ significantly between the visualization conditions. The whiskers indicate significant post-hoc comparisons (* denotes $p < 0.05$; ** denotes $p < 0.01$).}
    \label{fig:grip_matching_performance}
\end{figure}

The average grip-matching performance metrics for each visualization condition are presented in Figure~\ref{fig:grip_matching_performance}.

\paragraph{Mean Number of Matches}

%          Chisq   Df  Pr(>Chisq)  
% Zone     0.5834  1    0.44497  
% Vis      5.2909  1    0.02144 *
% Zone:Vis 5.8345  1    0.01571 *

% Zone = A:
%  contrast ratio     SE  df null z.ratio p.value
%  B / G    0.876 0.0348 Inf    1  -3.334  0.0009

% Zone = B:
%  contrast ratio     SE  df null z.ratio p.value
%  B / G    1.003 0.0398 Inf    1   0.082  0.9344

%   Vis   mean_dv sd_dv     n
% 1 B        5.73  1.98    28
% 2 G        6.13  2.36    28

As the mean number of matches was expected to be strictly positive and follow a right-skewed distribution we applied a Gamma distribution with a log link function.
Results showed that a significant main effect of Visualization ($\chi^2 = 5.29, p < 0.05$) with participants achieving more matches in the Gradient condition ($M = 6.13, SD = 2.36$) compared to the Bar condition ($M = 5.73, SD = 1.98$). There was also a significant Sensor-Region Mapping $\times$ Visualization interaction ($\chi^2 = 5.83, p < 0.05$), while the
main effect of sensor-region mapping was not significant ($\chi^2 = 0.58, p = 0.44$). Post-hoc tests revealed that the effect of visualization was significant only in Mapping A (Bar/Gradient ratio of means = $0.87$, $ p < 0.001$).

\paragraph{Mean Time to First Match}

%           Chisq Df   Pr(>Chisq)  
% Zone     2.9054  1    0.08828 .
% Vis      3.3232  1    0.06831 .
% Zone:Vis 2.8279  1    0.09264 .

%   Vis   mean_dv sd_dv     n
%   <chr>   <dbl> <dbl> <int>
% 1 B        4.00  1.14    28
% 2 G        3.69  1.02    28
%   Zone_Transformed mean_dv sd_dv     n
%   <chr>              <dbl> <dbl> <int>
% 1 A                   4.14 1.18     28
% 2 B                   3.55 0.903    28

As the block-level average time (in seconds) needed to make the first match in a trial would be strictly positively and right-skewed, we modeled it using a Gamma distribution and log link function.
Results showed no significant main effect of Sensor-Region Mapping ($\chi^2 = 2.90, p = 0.088$) or Visualization ($\chi^2 = 3.32, p = 0.068$), nor was there a significant interaction between the two ($\chi^2 = 2.82, p = 0.092$). The performance of the participants was similar across the Bar ($M = 4.00, SD = 1.14$) and Gradient ($M = 3.69, SD = 1.02$) visualization conditions. Similarly, their performance remained consistent between Mapping A ($M = 4.14, SD = 1.18$) and Mapping B ($M = 3.55, SD = 0.90$).
% Results showed no significant effects of Zone ($\chi^2 = 2.90, p=0.088$), Visualization ($\chi^2=3.32, p=0.068$) or their interaction ($\chi^2 = 2.82, p = 0.092$).

\paragraph{Average Match MAE}
% KS test failed with ordbeta regression, used gamma with log link function
%          Chisq  Df  Pr(>Chisq)  
% Zone     0.4226  1    0.51566  
% Vis      3.9774  1    0.04611 *
% Zone:Vis 0.0482  1    0.82615  
%   Vis   mean_dv  sd_dv     n
% 1 B       0.263 0.0443    28
% 2 G       0.256 0.0472    28

For mean MAE across all matches in a block, the ordered beta regression residuals did not satisfy model fit diagnostics despite the outcome being bounded between $0$ and $1$, so we fitted a GLMM with a Gamma family and log link function, which provided adequate fit. The mean MAE showed a significant main effect of Visualization ($\chi^2 = 3.98$, $p = .046$) with participants achieving lower MAE in the \textit{Gradient} condition ($M = 0.256, SD = 0.047$) compared to Bar condition ($M = 0.263, SD = 0.044$). The main effect of Sensor-Region Mapping ($\chi^2 = 0.42$, $p = .516$) and the Sensor-Region Mapping $\times$ Visualization interaction ($\chi^2 = 0.05$, $p = .826$) were not significant. These results indicate that Visualization condition influenced grip-matching accuracy consistently across both sensor-region mapping conditions, with the \textit{Gradient} visualization yielding lower absolute error.

%\paragraph{Mean Best Match MAE}
%
%           Chisq  Df  Pr(>Chisq)   
% Zone     0.5357  1   0.464211   
% Vis      8.4392  1   0.003672 **
% Zone:Vis 0.2070  1   0.649157  
%  Vis   mean_dv  sd_dv     n
%   <chr>   <dbl>  <dbl> <int>
% 1 B       0.239 0.0476    28
% 2 G       0.228 0.0510    28
% post-hoc
% Zone_Transformed = A:
%  contrast odds.ratio     SE  df null z.ratio p.value
%  B / G          1.07 0.0328 Inf    1   2.361  0.0182

% Zone_Transformed = B:
%  contrast odds.ratio     SE  df null z.ratio p.value
%  B / G          1.05 0.0317 Inf    1   1.753  0.0796

\paragraph{Mean Best Match MAE}
As the MAE of the best match was bounded between $0$ and $1$, we used ordered beta regression with a logit link. Results showed a significant main effect of Visualization ($\chi^2 = 8.44$, $p < 0.01$), with  errors being lower with the \textit{Gradient} visualization ($M = 0.228$, $SD = 0.051$) compared to the \textit{Bar} visualization ($M = 0.239$, $SD = 0.048$). Neither the main effect of sensor-region mapping ($\chi^2 = 0.53$, $p = 0.467$) nor the Sensor-Region Mapping $\times$ Visualization interaction ($\chi^2 = 0.20$, $p = 0.655$) was significant, indicating that the advantage of the \textit{Gradient} visualization was consistent across both sensor-region mappings.

% TLX
\subsection{Task Load \& Preference} \label{results:tlx}

NASA-TLX responses were analyzed using LMMs with sensor-region mapping and Visualization as fixed effects. No significant effects were observed for the overall weighted TLX score, mental demand, physical demand, frustration, or perceived performance. A significant main effect of sensor-region mapping was found for temporal demand ($\chi^2 = 5.99$, $p = .014$), with Mapping B participants reporting higher temporal demand ($M=55.9$, $SD=20.1$) than Mapping A participants ($M=37.9$, $SD=20.2$). For effort, a significant Sensor-Region Mapping $\times$ Visualization interaction was observed ($\chi^2=6.90,p=.009$ ); in the \textit{Bar} condition, Mapping A participants reported lower effort ($M=54.3$, $SD=21.4$) than Mapping B ($M=72.5$, $SD=13.6$, $p<.05$), whereas no significant difference between sensor-region mappings was found in the \textit{Gradient} condition ($p=.247$).

\paragraph{Visualization Preference} \label{sec:preference}
When asked to indicate their preferred visualization at the end of the study, participants were evenly split: 50\% (n=14) favored the Bar condition, while 50\% (n=14) favored the Gradient condition.

\subsection{Gaze Metrics} \label{results:gaze}
Gaze logs were processed to extract gaze positions per block using the recorded timestamps. From these, we computed gaze entropy, horizontal dispersion (x-dispersion), and vertical dispersion (y-dispersion) to characterize gaze behavior. Gaze entropy measures the unpredictability of gaze distribution across the visual scene, with lower entropy indicating more focused and systematic scanning~\cite{Bhavsar2017, ayala2023effects, LeeJung2025, Krejtz2015Gaze, Shiferaw2019Review, DiStasi2016}. Horizontal and vertical dispersion quantify the spatial spread of gaze along each axis, reflecting how broadly or narrowly a participant's attention was distributed across the feedback display. Raw gaze coordinates (in pixels) were converted to visual angle using the screen's physical dimensions and the participant's viewing distance, mapping screen positions to degrees of visual angle subtended at the eye. Dispersion was then computed as the standard deviation of visual angle values across all gaze samples within each trial, separately for the horizontal and vertical axes.

\paragraph{Gaze Entropy}

% Response: entropy
%            Chisq Df Pr(>Chisq)    
% Zone      0.6755  1    0.41115    
% Vis      25.9034  1  3.589e-07 ***
% Zone:Vis  4.7390  1    0.02949 *  

% Zone = A:
%  contrast ratio     SE  df null z.ratio p.value
%  B / G     1.08 0.0165 Inf    1   5.162 <0.0001

% Zone = B:
%  contrast ratio     SE  df null z.ratio p.value
%  B / G     1.03 0.0161 Inf    1   1.998  0.0457

%   Vis   mean_dv sd_dv     n
% 1 B        6.44 0.712    26
% 2 G        6.09 0.744    26

Gaze entropy showed a significant main effect of visualization type ($\chi^2 = 25.90$, $p < .001$), with greater gaze entropy in the \textit{Bar} condition ($M = 6.44$ bits, $SD = 0.71$) compared to the \textit{Gradient} ($M = 6.09$ bits, $SD = 0.74$). The main effect of Sensor-Region Mapping was not significant ($\chi^2 = 0.68$, $p = .411$), but the Sensor-Region Mapping $\times$ Visualization interaction was significant ($\chi^2 = 4.74$, $p = .029$). Follow-up pairwise comparisons showed that the Bar-vs-Gradient difference in entropy was significant in both mapping conditions, but substantially larger in Mapping A (ratio $= 1.08$, $z = 5.16$, $p < .0001$) than in Mapping B (ratio $= 1.03$, $z = 2.00$, $p = .046$), indicating that the visualization effect on gaze entropy was attenuated under Mapping B relative to Mapping A.

\paragraph{Gaze X-dispersion}

% Response: raw_x_dispersion
%            Chisq Df Pr(>Chisq)    
% Zone      0.0067  1     0.9349    
% Vis      33.2738  1  8.005e-09 ***
% Zone:Vis  0.1127  1     0.7371  

%   Vis   mean_dv sd_dv     n
% 1 B        4.65  1.07    26
% 2 G        5.23  1.33    26

Gaze X-dispersion showed a significant main effect of visualization ($\chi^2 = 33.27, p < .001$), indicating that participants scanned more broadly from left to right with the Gradient visualization ($M=5.3^\circ$, $SD = 1.07^\circ$) compared to the Bar ($M=4.65^\circ$, $SD = 1.33^\circ$) across both sensor-region mapping conditions. Neither the main effect of Sensor-Region Mapping ($\chi^2 = 0.007, p = .93$) nor the Sensor-Region Mapping $\times$ Visualization interaction ($\chi^2 = 0.11, p = .73$) were significant.

\paragraph{Gaze Y-dispersion}

% Response: raw_y_dispersion
%            Chisq Df Pr(>Chisq)    
% Zone      0.1051  1  0.7457692    
% Vis      10.8995  1  0.0009619 ***
% Zone:Vis  1.5039  1  0.2200780  
% had to use GLMM with gaussian and log link as the LMM fit residuals violated KS test/normality
% 1 B        10.4  2.64    26
% 2 G        11.1  3.25    26
Gaze Y-dispersion was analyzed using a GLMM with a Gaussian distribution and log link function.

Results showed a significant main effect of Visualization ($\chi^2 = 10.89, p < .001$), with participants exhibiting significantly lesser vertical gaze coverage in the \textit{Bar} visualization condition ($M=10.4^\circ$, $SD=2.64^\circ$) compared to the \textit{Gradient} ($M=11.1^\circ$, $SD=3.25^\circ$) across both the sensor-region mapping conditions. Neither the main effect of Sensor-Region Mapping ($\chi^2 = 0.10, p = 0.74$) nor the Sensor-Region Mapping $\times$ Visualization interaction ($\chi^2 = 1.50, p = 0.22$) were significant, indicating that the effect of Visualization condition on vertical gaze dispersion was consistent regardless of sensor-region mapping.

\begin{figure}[ht]
    \centering
    \subfloat[Gaze Entropy\label{fig:gaze_entropy}]{%
        \includegraphics[width=0.32\linewidth]{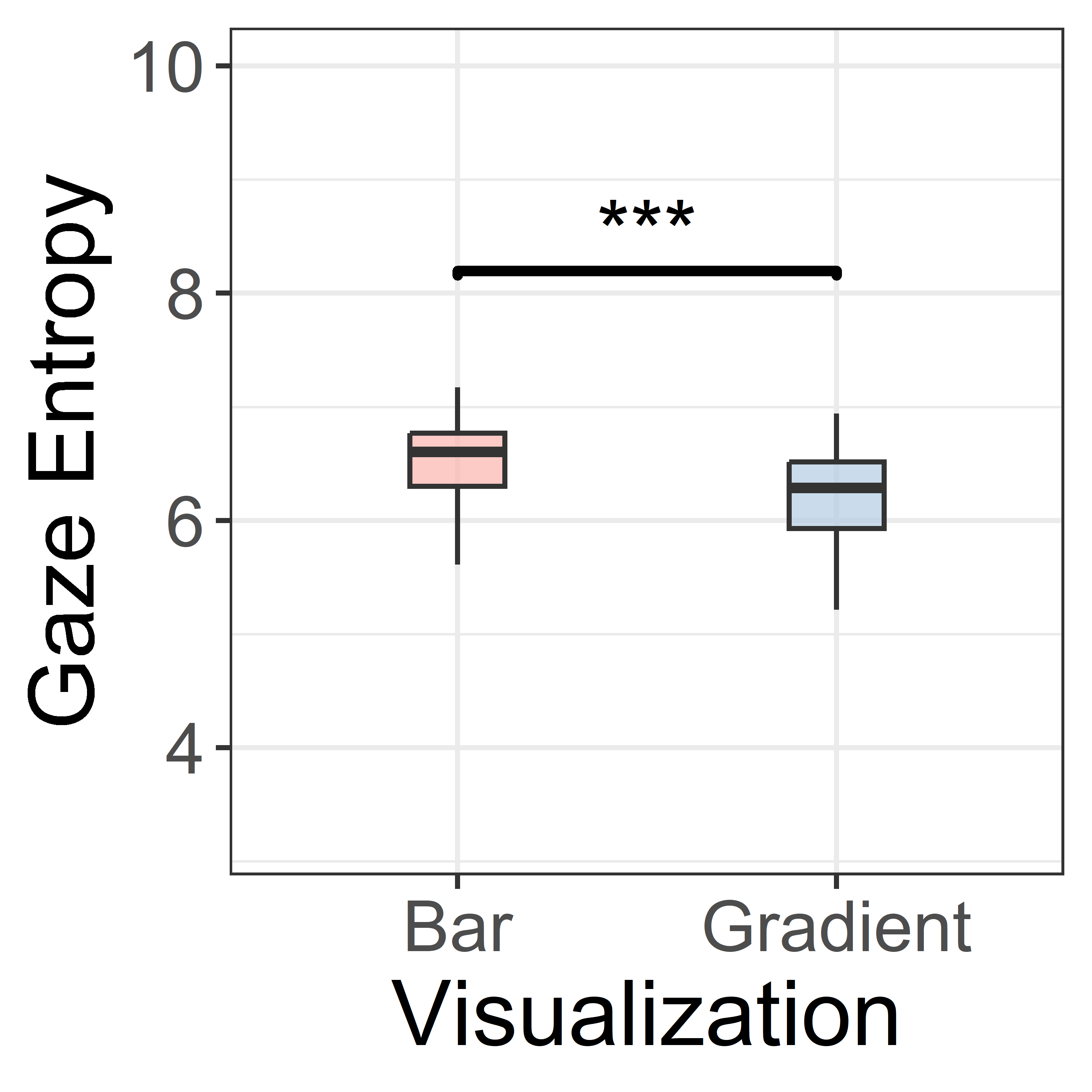}}
    \hfil
    \subfloat[Gaze X\nobreakdash-dispersion\label{fig:x_dispersion}]{
        \includegraphics[width=0.32\linewidth]{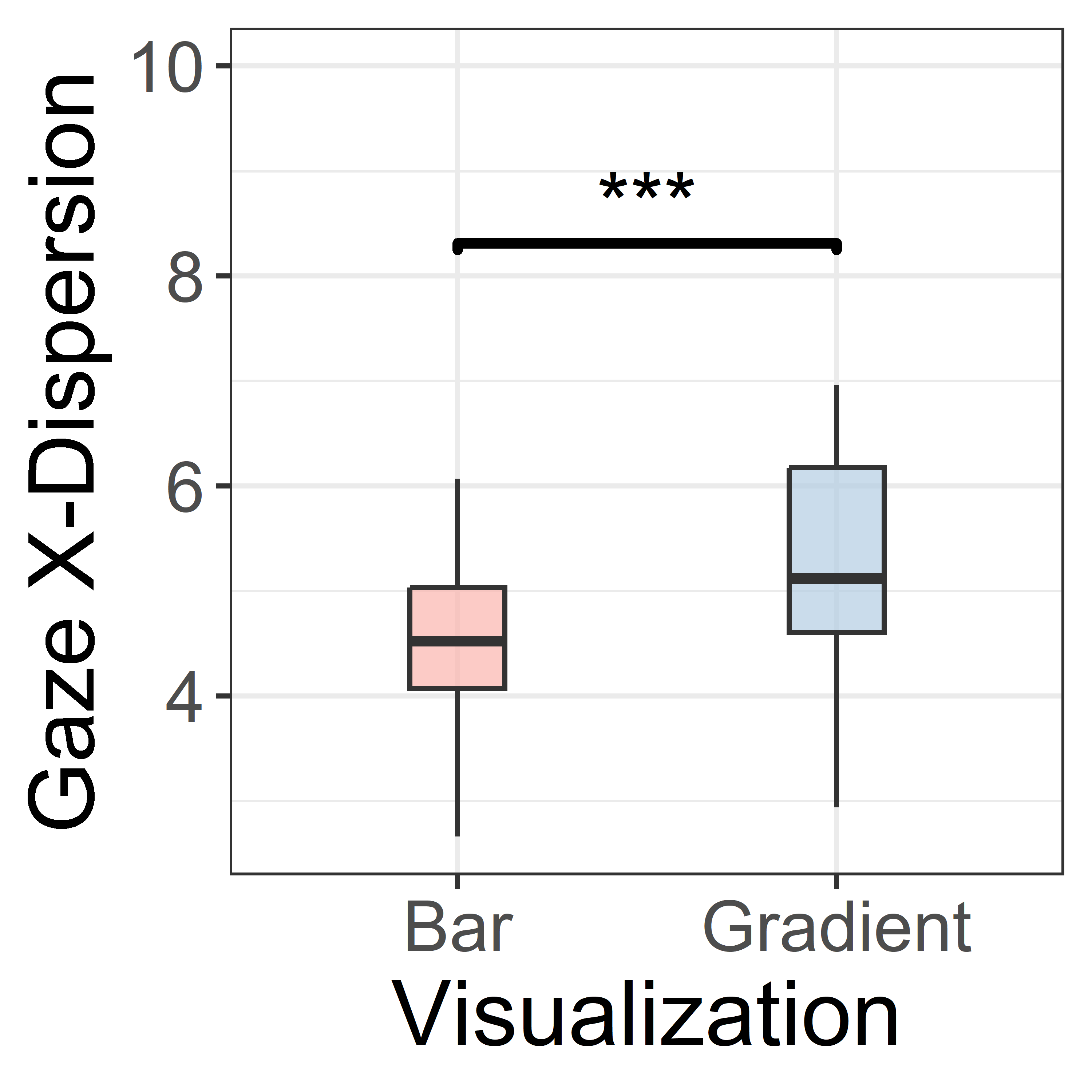}}
    \hfil
    \subfloat[Gaze Y\nobreakdash-dispersion\label{fig:y_dispersion}]{
        \includegraphics[width=0.32\linewidth]{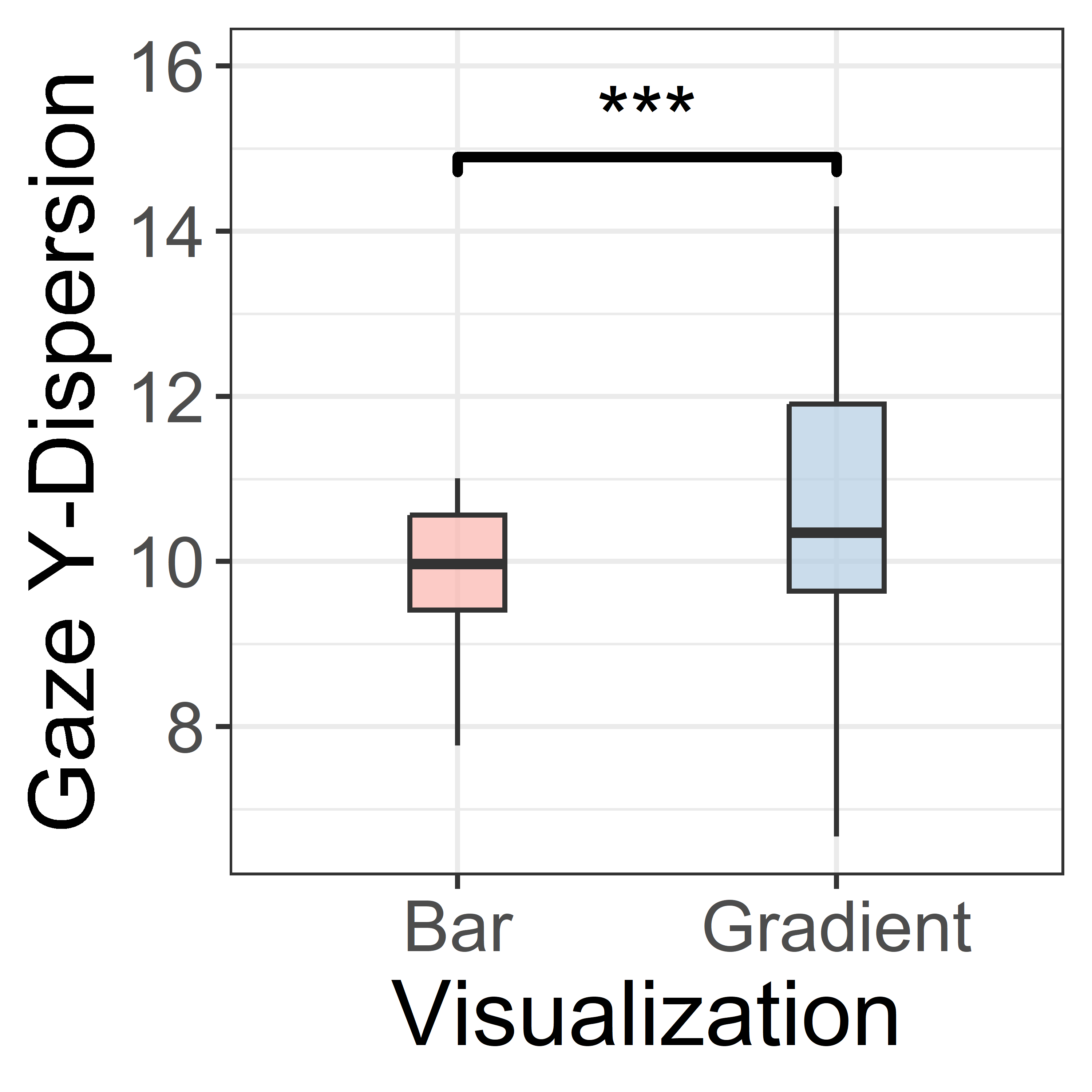}}
    \caption{The gaze behavior of the participants is measured using mean (a) gaze entropy, (b) gaze X-dispersion and (c) gaze Y-dispersion over a trial block. Participants exhibited significantly higher gaze entropy with the \textit{Bar} visualization, indicating more exploratory and less structured scanning patterns. Both gaze X-dispersion and Y-dispersion were significantly higher with the \textit{Gradient} visualization, reflecting broader visual scanning across the interface. The whiskers indicate significant post-hoc comparisons (*** denotes $p<0.001$).}
    \label{fig:gaze_metrics}
\end{figure}

\section{Qualitative Results}

Approximately five hours of post-study interviews with the 28 participants were audio recorded and transcribed. We analyzed these transcripts together with participants' open-ended survey responses. Two researchers independently read a subset of the transcripts and generated initial codes inductively, then met to consolidate these into a shared codebook. Both researchers then applied the codebook across the full set of transcripts and open-ended responses, meeting to discuss and resolve disagreements and to refine code definitions as needed. Related codes were grouped into the topics reported below. Where we report participant counts, these are intended to convey how widely a view was expressed among our participants rather than to support statistical inference.

\subsubsection{``How Much'' Versus ``Where''} \label{qual:howmuchwhere}
Participants consistently described the two encodings as answering different questions: the Bar told them \emph{how much} pressure to apply, while the Gradient told them \emph{where} to apply it.
 
Ten participants identified the Bar condition as more effective for conveying precise levels of pressure intensity compared to the Gradient condition. P77 (Mapping A) noted that the change in bar fill with applied pressure made it ``the most intuitive of all'' and indicated that the numerical and color cues offered supplementary information. The value of different cues was also emphasized by P31 (Mapping A), who stated, ``higher the bar goes up the color also changes, which tells me I'm applying too much pressure.'' Three participants similarly found the numerical cues helpful to correlate the discrete pressure ranges to the applied force. In contrast, four participants perceived the Gradient condition as less precise for matching pressure intensity, with P15 (Mapping A) noting, ``The gradient tells you more what to grip with less precision on how hard to grip.''

Conversely, participants generally found the Gradient visualization more effective for mapping pressure to specific hand locations. P13 and P39 noted that the continuous shading enabled them to identify the regions that needed adjustments, and it supported targeted muscle flexing by guiding them where to apply or avoid pressure (P33, P45, P76, P77). P76 said the differences in the highlighted areas indicated to them that ``they shouldn't use that muscle to touch the object,'' illustrating how the color gradient supported the development of mental models for grip placement. Several participants attributed the ease of use of the Gradient condition to its spatial layout, with P14 (Mapping B) explaining it was ``easier to understand because all [the] fingers are lining up.'' Some participants reported that the Bar visualization initially presented spatial ambiguity, as it omitted linking each bar to a specific finger or region (P12). However, once participants developed familiarity, five reported that the Bar's simpler structure reduced monitoring effort. P29 explained that ``it was easier to match the bars on the four regions rather than gradients.''
 
Although the Bar and Gradient conditions visualize the same aggregated data, participants described them as conveying the answers to \emph{where} and \emph{how much} differently. According to them, the Gradient spread attention across the hand, while the Bar's exact numerical targets imposed a translational effort. Rather than judging one encoding as better overall, participants assigned each a role, which is consistent with the even split in stated preference reported in Section~\ref{sec:preference}: precise magnitude guidance for the Bar and spatial guidance for the Gradient.

\subsubsection{Perceived Stability and Noise} \label{qual:stability}
Participant perception of feedback stability varied between the two visualization conditions. The Gradient was perceived as being ``noisy'' and harder to read precisely (P22, P77), which made it harder for some to notice their ``small mistakes,'' while others found it allowed for a ``more smooth and satisfying experience'' (P37, Mapping A). The Bar visualization was perceived as more stable and responsive to overall grip strength by seven participants, %(P16, P17, P18, P29, P32, P75, P77), 
but its numerical display was described as ``oscillating'' (P14, Mapping B) and frustrating because it was difficult to maintain the exact target values (P4, P14, P18, P37).

\subsubsection{Attentional and Physical Demand} \label{qual:demand}
Participants reported different \emph{kinds} of demand for the two encodings. For some, the spatial detail of the Gradient condition translated into broader monitoring demand. They reported needing to track ``too many things'' (P77, Mapping A), the feedback having ``a bit more noise'' (P22, Mapping B), and requiring focus on ``almost every corner'' of the hand (P26, Mapping B). The Bar's limited visual elements, while lower in granularity, reduced the number of cues these participants needed to monitor simultaneously, making it ``more cognitively manageable'' (P16, Mapping B).

Others located the demand in the Bar's numerical targets instead. These participants described the Gradient condition as requiring ``less mental effort'' (P11), feeling ``less taxing and easier'' (P41, Mapping A), and producing fewer feelings of frustration with less ``mental load'' (P39, Mapping A). By contrast, the Bar visualization was considered more ``physically strenuous and effortful'' (P14, Mapping B) and to require ``more attention'' (P41, Mapping A), with P41 mentioning that the ``bars felt really specific and honestly unattainable.'' Taken together, these accounts suggest that the Gradient distributed attention more widely across the hand, whereas the Bar concentrated attention on a smaller display but added the work of translating numerical targets into physical grip adjustments.

\subsubsection{Strategies for Managing Multiple Regions Under Time Pressure}\label{qual:strategies}
Participants described strategies that reduced the matching task to something they could manage within the 20-second window. Ten participants %(P11, P14, P17, P26, P29, P36, P38, P41, P71, P76) 
incorporated the visualization sequentially and made adjustments to the grip one finger or region at a time. P29 (Mapping A) described this strategy: ``I focused on one area or a specific region\ldots then I figured out if I was pressing too hard or too little and then I tried to adjust\ldots while trying not to affect other regions too much.'' Six participants %(P14, P20, P34, P39, P41, P77) 
also indicated that they prioritized functionally critical areas, particularly the thumb, index, and middle fingers, instead of monitoring the entire hand.
 
Eleven participants often settled on a ``close enough'' match rather than trying to achieve exact alignment. Three specifically mentioned they stopped when the visualization appeared sufficiently similar to the reference, with P15 (Mapping A) explaining that trying to improve the match might cause them to ``mess up somewhere else.'' Others stopped when matches looked ``close'' or ``near there'' under time pressure (P17, P33, P34). Colors were used as the primary cue to determine if there was a match. P41 (Mapping A) mentioned that they were ``looking at the number for guidance'' but for the judgment they were ``just looking at the color.'' Five participants also captured mental snapshots of target sensor-regions before adjusting their grip.
 
P15's concern about ``mess[ing] up somewhere else'' shows that participants understood the hand's regions to be coupled and deliberately limited how much of the feedback they acted on at once, so as to avoid disturbing a partial match already achieved. These accounts share a common function of trading completeness for control, revealing a tradeoff between precision and the heuristic shortcuts used to achieve the best performance possible under time pressure.

\subsubsection{Shifting Reliance on the Display} \label{qual:reliance}
Participants described their reliance on the feedback changing over the session, from deliberate reading toward adjustments made without conscious reference to the display. Five participants suggested that the tacit knowledge could be communicated more effectively through a staged process. P32 explained that ``the bar graph would be easier to start off just to get a sense and feel for how it works with less information,'' while P71 described the Bar condition as ``not so overwhelming'' and ``accessible.''
 
Familiarity also influenced how participants incorporated the feedback and made adjustments, with many shifting from sequential adjustments toward faster, holistic matching strategies. Many participants (P4, P5, P11, P16, P18, P26, P41, P75) mentioned making instinctive adjustments without conscious deliberation. P18 (Mapping B) mentioned that the adjustments ``felt like intuition'' and P4 (Mapping B) stated they understood how to ``alter [their] adjustments based on the visualization.''
 
Six participants described developing an intuitive sense of hand positioning and force application that they reported was absent before using the system. Participants P13 and P18, who had no prior sports experience, reported recognizing and understanding proper gripping techniques for the first time. That participants with no prior sports experience reported recognizing correct grips for the first time suggests the feedback did more than support in-the-moment matching: as participants became more familiar, they described an emerging sense of pressure and placement that they applied with less direct reference to the display.

\section{Discussion and Design Implications}

Our quantitative and qualitative results suggest that the choice between continuous (\textit{Gradient}) and discrete (\textit{Bar}) contact force visualizations, along with their underlying sensor-region mapping, may have shaped how participants processed feedback, corrected errors, and formulated motor strategies.

\subsection{Continuous visual encodings help with accurate corrections.}

Participants matched the reference grip more accurately with the \textit{Gradient} condition, yielding lower MAE (\textbf{RQ1}, Section \ref{results:matching}). Both encodings were driven by the same 20 sensors aggregated into the same $K=4$ sensor-regions, so this advantage does not follow from one display covering more of the hand than the other, but from how those values are presented. By overlaying feedback across the hand, the continuous spatial mapping may have minimized the transformation costs required to map visual error signals onto specific hand regions~\cite{larsen1998effects}, which may have led to lower MAE. The spatial alignment in the \textit{Gradient} condition likely facilitated faster development of internal models of the pressure application, allowing for precise grip corrections~\cite{wolpert1995internal, wolpert2011principles}. By rendering feedback on the region it concerns, the \textit{Gradient} condition aligns with the proximity compatibility principle, lowering the cost of mapping pressure magnitude to hand location \cite{WickensCarswell1995} (Section~\ref{qual:howmuchwhere}). The continuous field may also have supported holistic processing of the whole-hand pressure pattern~\cite{Ariely2001, AlvarezOliva2009, Whitney2018} consistent with the lower gaze entropy observed in this condition. Conversely, the aggregated targets in the \textit{Bar} condition provided a simplified error signal~\cite{McDougle2016Taking}. While sufficient for global force scaling, this abstraction lacked the topographical detail necessary for the fine-grained adjustments required to produce accurate matches.

Designers of grip pressure feedback systems should favor continuous spatial visualizations when grip precision is the primary objective \textbf{(RQ1)}. By preserving the anatomical topography of the hand, continuous encodings minimize the cognitive translation between visual error and motor action, enabling participants to pinpoint adjustments intuitively.

\subsection{Visual encodings shape attention and correction strategies.}
The \textit{Bar} condition's reduced spatial fidelity affected matching throughput, with participants completing fewer matches than in the \textit{Gradient} condition, though this effect was significant only under the grip-tailored sensor-region mapping (Mapping A) (Section~\ref{results:matching})~\textbf{(RQ1)}. While time to first match did not differ significantly between visual encodings, participants described a slower, analytical ``finger-by-finger'' adjustment strategy for managing the \textit{Bar}'s discrete regions under time pressure (Section~\ref{qual:strategies}), and separately characterized the \textit{Bar} condition as ``physically strenuous'' and attentionally demanding, locating that demand in the specificity of the targets themselves (Section~\ref{qual:demand}). This sequential correction pattern may have contributed to the reduced matching rate~\cite{wolpert1995internal, wolpert2011principles, Sigrist2013a}. However, not all participants found the \textit{Bar} condition more demanding. Some participants found its smaller set of cues more cognitively manageable and reported that its simpler structure reduced monitoring effort once they were familiar with it (Section~\ref{qual:howmuchwhere}), while others described the \textit{Gradient} condition as requiring attention across the whole hand (Section~\ref{qual:demand}). This split, together with the absence of a significant main effect of visualization on any NASA-TLX subscale \textbf{(RQ1)}, suggests the encodings differ in the \textit{kind} of demand imposed rather than its total amount. The \textit{Gradient} condition distributed monitoring across the hand, whereas the \textit{Bar} condition concentrated it on fewer elements while adding the work of converting numerical targets into grip adjustments.

Gaze entropy and dispersion, which are proxies for visual attention ~\cite{LeeJung2025, ayala2023effects, Bhavsar2017}, show a parallel pattern~\textbf{(RQ1)}. Entropy was significantly higher in \textit{Bar} (Section~\ref{results:gaze}), consistent with more frequent attention-switching as participants translated numerical values into hand regions~\cite{Bhavsar2017, ayala2023effects, LeeJung2025}. This held under both mappings but was attenuated under the generic one~\textbf{(RQ2)}, mirroring the effort interaction above. Dispersion moved oppositely as gaze was wider in \textit{Gradient} on both axes, reflecting that encoding's whole-hand signal rather than added search. Because entropy was computed over fixed screen-space bins, it would ordinarily rise with the area gaze covers. \textit{Gradient}'s combination of wider dispersion and lower entropy indicates a few widely spaced fixations rather than dense region-by-region sampling suggesting that participants read the pattern at a glance. \textit{Bar}'s lack of an explicit bar-to-region link (Section~\ref{qual:howmuchwhere}) may explain this pattern, as participants would have needed to sample the target profile repeatedly to recover that correspondence. In short, \textit{Bar}'s narrower gaze footprint was not a more efficient one as it paired with higher entropy and higher matching error, with the latter being consistent across both mappings.

Together, these findings suggest that designers should consider the amount of information shown and the attentional strategy an encoding elicits \textbf{(RQ1)}. Continuous encodings may be particularly useful when users must make repeated corrections under time pressure. When discrete encodings are necessary, explicit visual links between each bar and the hand region it summarizes may reduce the correspondence-search cost reflected in \textit{Bar} condition's gaze entropy.

\subsection{Perceived magnitude precision did not track matching accuracy.}

Many participants judged the \textit{Bar} more precise for reading pressure magnitude (Section~\ref{qual:howmuchwhere}) even though they matched more accurately, completed more matches (Section~\ref{results:matching}), and exhibited lower gaze entropy (Section~\ref{results:gaze}) with the \textit{Gradient}~\textbf{(RQ1, RQ3)}. In the interviews, many participants found the \textit{Bar} more intuitive and accessible, and several described the \textit{Gradient} as less precise for judging how hard to press (Section~\ref{qual:howmuchwhere}). This is consistent with established findings that position and length are among the most accurate channels for reading a specific value \cite{ClevelandMcGill1984}, so the \textit{Bar} was better suited to judging how much pressure to apply. But reading an exact value is not what the task required. MAE averages error across all active sensors  (Eq.~\ref{eq:mae}), so matching rewards reproducing a pressure distribution across the hand rather than hitting a target magnitude in any one region. The encoding participants judged more precise was therefore not the one they matched more accurately with \textbf{(RQ3)}. Neither encoding was perceived as steady: participants described the \textit{Gradient} condition as noisy and the \textit{Bar} condition's numbers oscillating and hard to hold at the target (Section~\ref{qual:stability}). 

The difference is that the \textit{Bar} named a value to reach through the discrete numbers on the display, allowing participants to interpret fluctuations as a gap to close. The \textit{Gradient} provided a less explicit reference against which to judge such fluctuations. Therefore, participants may have judged precision by whether the display stated a target, not by how closely it let them match one \textbf{(RQ3)}. On this reading, the two encodings suit different stages of learning, starting with the \textit{Bar} condition followed by the \textit{Gradient} condition after familiarization, a sequence participants proposed themselves (Section~\ref{qual:reliance}).

We recommend that designers should leverage discrete visualizations for initial familiarization, transitioning to continuous spatial feedback as learners progress toward precise closed-loop correction~\textbf{(RQ3)}.

\subsection{Mapping alignment shapes the effectiveness of visual feedback.}

Beyond visual encoding, our results suggest that the effectiveness of contact force feedback depends on whether the sensor-region mapping strategy aligns with the biomechanical demands of the grip \textbf{(RQ2)}. That dependence, however, was one of cost and not of capability as mapping affected reported workload (Section~\ref{results:tlx}) but had no main effect on matching error or throughput (Section~\ref{results:matching}). Mapping also had no main effect on gaze entropy or on gaze dispersion along either axis (\textbf{RQ2}, Section~\ref{results:gaze}). The grip-tailored mapping (Mapping~A) concentrated feedback on the contact regions most relevant to each grip type, with the regions defined by the \textit{virtual fingers} in the GRASP taxonomy~\cite{Feix2016}. In contrast, the non-tailored mapping (Mapping B) distributed sensors across less grip-relevant regions, introducing spatial ambiguity that may have increased the interpretive burden on the user \cite{Sweller2010, Sigrist2013a}. Because each visualization summarized pressure within regions defined by the sensor-region mapping, visualization and mapping were not independent design choices.

The higher temporal demand reported by Mapping~B participants across both visualization conditions (\textbf{RQ2}, Section~\ref{results:tlx}) suggests that resolving which hand regions required adjustment consumed time, effort, and attention that would otherwise have gone to grip correction. To find a match within the time constraints of a trial, participants described a strategy of working through the hand one region at a time and settling for a close-enough match, which may explain the higher temporal demand (Section~\ref{qual:strategies}).

The two significant Sensor-Region Mapping $\times$ Visualization interactions contextualize that cost (Sections~\ref{results:matching}). The \textit{Gradient} condition's advantage in number of matches held only under Mapping~A, and the effort cost of Mapping~B appeared only under the \textit{Bar} condition. One reading is that spatial alignment operates at two coupled levels: how sensors are aggregated into regions and how summaries of those regions are represented visually. The \textit{Gradient} condition can only make visible the task-relevant structure created by the mapping; when the sensor-region mapping did not follow the grip, its throughput advantage disappeared. Because the \textit{Bar} condition did not explicitly show where each summarized region lay on the hand, the same mapping mismatch may instead have surfaced as additional translational effort. Mapping did not affect reported effort in the \textit{Gradient} condition, which suggests a continuous display can absorb some of the cost of a sensor-region mapping that does not match the grip. Participants under Mapping~B still described the \textit{Gradient} condition as easier to understand because the fingers lined up (Section~\ref{qual:howmuchwhere}), indicating that correspondence to hand layout remained legible even where the sensor assignment did not follow the grip. Feedback actionability therefore depends not only on how information is displayed but also on how well its spatial structure maps onto the relevant action surfaces~\cite{Feix2016, Sundaram2019LearningSignatures, nargund2026embedded}.

When generating pressure sensor-region mappings for the hand, designers should pair continuous spatial encodings with mapping strategies grounded in established grasp taxonomies or task-specific biomechanical analyses rather than applying uniform or anatomically generic mappings \textbf{(RQ2)}.

\section{Limitations}

The current prototype was designed exclusively for right-handed users and built in a single fixed form factor, limiting the generalizability of the findings.

The system also relies on FSR sensors, which are susceptible to erratic readings and electromagnetic noise at high resistance values. These signal artifacts introduce extraneous cognitive load that higher-fidelity sensing hardware could potentially mitigate \cite{Sweller2010, McDougle2016Taking}.

An off-the-shelf cotton glove was used, and its fit was not tailored to individual participants. We did not characterize the fabric's stiffness, and variation in stiffness, fit, and hand morphology (e.g., hand length versus breadth) across participants may have shifted sensor positions relative to anatomical landmarks such as the abductor \textit{pollicis brevis}. These differences affect how pressure transferred to the sensors and likely contributed error to the MAE in Eq.~\ref{eq:mae}. The reference profiles were recorded using the same off-the-shelf glove and carry the same sensor-placement variability. Differences in hand morphology between the reference experts and our participants may therefore contribute to the MAE in Eq.~\ref{eq:mae} independently of how accurately a participant reproduced the intended grip.

The experimental tasks involved static grip adjustments on physical sports objects in a laboratory setting. Real-world sports contexts require these grips to be executed and maintained during high-velocity, dynamic movements (e.g., hitting a ball while swinging a table tennis paddle). The current study does not account for the additional biomechanical forces and complex attentional demands present in fully dynamic environments \cite{Endsley95, wickens2008multiple}.

Finally, since the mappings were defined based on grip type, which we evaluated across the two sports, a sport–mapping interaction is expected. Sport was therefore treated as task context rather than as an experimental factor, so our analyses collapse across it and we do not report sport-specific effects.

\section{Future Work}

While our study demonstrates that real-time visual feedback can successfully externalize tacit grip knowledge, long-term motor skill transfer should be investigated by removing the real-time feedback and testing whether tacit grip skills are retained \cite{Salmoni1984, Sigrist2013b}.

Furthermore, our qualitative results also point to a staged motor guidance approach. Future designs should explore adaptive visualization pipelines that dynamically adjust whether discrete or continuous visual encodings are presented based on real-time user performance \cite{Fitts1967Human, guadagnoli2004challenge}. For instance, systems could automatically transition from the accessible, discrete \textit{Bar} condition to the continuous \textit{Gradient} condition once the user's grip-matching MAE stabilizes below a predetermined threshold \cite{Sherwood1988, Sweller2010}.

Lastly, to improve ecological validity, systems should be assessed in real-world settings by evaluating grip performance during dynamic, high-velocity scenarios, such as actively pitching a baseball or swinging a table tennis paddle. However, making this transition will require future hardware iterations to be significantly more robust, incorporating stretchable and durable materials, noise-resistant sensor arrays, and support for a diverse range of hand sizes and handedness \cite{Sundaram2019LearningSignatures}.

\section{Conclusion}

Tacit grip knowledge, an intuitive yet often invisible and embodied understanding of how to apply hand pressure, remains difficult to articulate or teach. In this work, we investigated how to  externalize this invisible data through real-time visual feedback to support hand-grip guidance. We evaluated whether concurrent visual feedback of hand grip pressures should be visualized using continuous or discrete encodings~\textbf{(RQ1)}, with our results indicating a precision-usability tradeoff: the continuous \textit{Gradient} condition yielded more accurate matching and lower gaze entropy without a measurable change in reported workload, yet many participants found the discrete \textit{Bar} condition more intuitive for initial familiarization~\textbf{(RQ3)}. Furthermore, grip-tailored sensor-region mappings significantly reduced perceived temporal demand and effort~\textbf{(RQ2)}, likely by reducing the spatial ambiguity participants faced when interpreting which hand regions required adjustment. Based on these findings, we propose design guidelines for haptic-to-visual feedback systems that externalize grip pressure.

\section*{Acknowledgments}
We thank Erika Hucal for her help constructing the instrumented glove prototype.

\bibliographystyle{IEEEtran}
\bibliography{works}

\newpage

\begin{IEEEbiography}[{\includegraphics[width=1in,height=1.25in,clip,keepaspectratio]{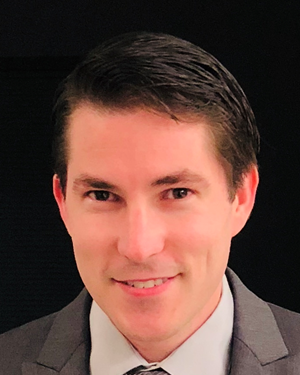}}]{Andrew L. Huard} (M'08) received the B.S. and M.S. degrees in electrical engineering from California Polytechnic State University, San Luis Obispo, CA, USA, in 2010. He is currently pursuing the Ph.D. degree in electrical and computer engineering at the University of California, Santa Barbara, CA, USA. His research interests include developing information theory frameworks for human-computer interaction, electromagnetic propagation, and electromagnetic interference. He is a Senior Principal Electrical Engineer with Raytheon Intelligence and Space, Electronic Warfare Systems, in Goleta, CA, USA. He holds two patents in passive microwave direction finding.
\end{IEEEbiography}

\begin{IEEEbiography}[{\includegraphics[width=1in,height=1.25in,clip,keepaspectratio]{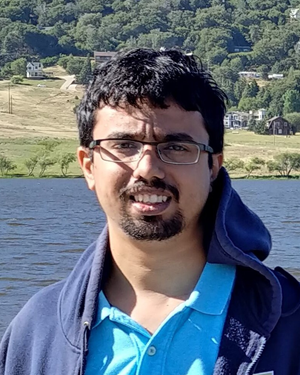}}]{Avinash Ajit Nargund} (M'24) received the B.S. degree in telecommunication engineering from R.V. College of Engineering, India, in 2015, and the M.S. degree in electrical and computer engineering from Georgia Institute of Technology, Atlanta, GA, USA in 2018. He is currently pursuing the Ph.D. degree in electrical and computer engineering at the University of California, Santa Barbara, CA, USA. His research interests include developing information theory frameworks for human-computer interaction and radar for autonomous vehicle engineering.
\end{IEEEbiography}

\begin{IEEEbiography}[{\includegraphics[width=1in,height=1.25in,clip,keepaspectratio]{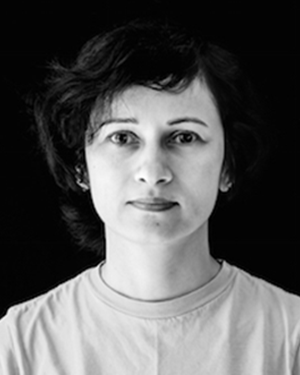}}]{Dr. Misha Sra} received the S.M. degree in media arts and sciences from the MIT Media Lab, Cambridge, MA, USA, in 2013 and the Ph.D. degree in media arts and sciences from the MIT Media Lab in 2018. She is currently the Gerngross Assistant Professor in the Department of Computer Science at the University of California, Santa Barbara, CA, USA. Her research focuses on human-computer interaction, extended reality, and human-centered artificial intelligence (AI), with an emphasis on designing novel interfaces, interactions, tools, and systems for psychomotor skill acquisition, task guidance, and creative collaboration. Her work explores the implications of AI-infused experiential interfaces, human-AI interaction, and the shift toward human-centric AI systems to augment human capabilities. Dr. Sra’s awards and honors include a Faculty Fellowship from the Hellman Foundation, an Early CAREER Award from the National Science Foundation, a Google Research Scholar Award, the Best Dissertation Award from the IEEE Visualization and Graphics Technical Committee, a Best Paper Award from ACM CHI, a Silver Edison Award, and a Best Paper Award from ACM VRST.
\end{IEEEbiography}

% your own custom macro containing the $\backslash${\tt{includegraphics}} command to make things
 

%\bf{If you include a photo:}\vspace{-33pt}
%\begin{IEEEbiography}[{\includegraphics[width=1in,height=1.25in,clip,keepaspectratio]{Template_TVCG/fig1.png}}]{Michael Shell}
%Use $\backslash${\tt{begin\{IEEEbiography\}}} and then for the 1st argument use $\backslash${\tt{includegraphics}} to declare and link the author photo.
%Use the author name as the 3rd argument followed by the biography text.
%\end{IEEEbiography}



%\vfill

\end{document}